\documentclass[12pt]{article}
\usepackage[utf8]{inputenc}
\usepackage[english]{babel}
\usepackage{a4wide}
\usepackage{amsfonts}
\usepackage{amssymb}
\usepackage{graphics,amsmath}
\usepackage{graphicx}
\usepackage{cancel}
\usepackage{cite}
\usepackage{mathrsfs}
\usepackage[usenames,dvipsnames,svgnames,table]{xcolor}
\usepackage{slashed}
\usepackage{etoolbox}

\DeclareMathOperator*\res{res}

\def\hybrid{\topmargin -20pt    \oddsidemargin 0pt
        \headheight 0pt \headsep 0pt
        \textwidth 6.35in       
        \textheight 9.25in       
        \marginparwidth .875in
        \parskip 5pt plus 1pt   \jot = 1.5ex}

\hybrid

\def\baselinestretch{1.2}

\catcode`\@=11

\def\marginnote#1{}
\newcount\hour
\newcount\minute
\newtoks\amorpm
\hour=\time\divide\hour by60
\minute=\time{\multiply\hour by60 \global\advance\minute by-\hour}
\edef\standardtime{{\ifnum\hour<12 \global\amorpm={am}%
        \else\global\amorpm={pm}\advance\hour by-12 \fi
        \ifnum\hour=0 \hour=12 \fi
        \number\hour:\ifnum\minute<10 0\fi\number\minute\the\amorpm}}
\edef\militarytime{\number\hour:\ifnum\minute<10 0\fi\number\minute}

\def\draftlabel#1{{\@bsphack\if@filesw {\let\thepage\relax
   \xdef\@gtempa{\write\@auxout{\string
      \newlabel{#1}{{\@currentlabel}{\thepage}}}}}\@gtempa
   \if@nobreak \ifvmode\nobreak\fi\fi\fi\@esphack}
        \gdef\@eqnlabel{#1}}
\def\@eqnlabel{}
\def\@vacuum{}
\def\draftmarginnote#1{\marginpar{\raggedright\scriptsize\tt#1}}

\def\draft{\oddsidemargin -.5truein
        \def\@oddfoot{\sl preliminary draft \hfil
        \rm\thepage\hfil\sl\today\quad\militarytime}
        \let\@evenfoot\@oddfoot \overfullrule 3pt
        \let\label=\draftlabel
        \let\marginnote=\draftmarginnote
   \def\@eqnnum{(\theequation)\rlap{\kern\marginparsep\tt\@eqnlabel}%
\global\let\@eqnlabel\@vacuum}  }

\def\preprint{\twocolumn\sloppy\flushbottom\parindent 2em
        \leftmargini 2em\leftmarginv .5em\leftmarginvi .5em
        \oddsidemargin -.5in    \evensidemargin -.5in
        \columnsep .4in \footheight 0pt
        \textwidth 10.in        \topmargin  -.4in
        \headheight 12pt \topskip .4in
        \textheight 6.9in \footskip 0pt
        \def\@oddhead{\thepage\hfil\addtocounter{page}{1}\thepage}
        \let\@evenhead\@oddhead \def\@oddfoot{} \def\@evenfoot{} }

\def\numberbysection{\@addtoreset{equation}{section}
        \def\theequation{\thesection.\arabic{equation}}}

\def\underline#1{\relax\ifmmode\@@underline#1\else
        $\@@underline{\hbox{#1}}$\relax\fi}

\def\titlepage{\@restonecolfalse\if@twocolumn\@restonecoltrue\onecolumn
     \else \newpage \fi \thispagestyle{empty}\c@page\z@
        \def\thefootnote{\fnsymbol{footnote}} }

\def\endtitlepage{\if@restonecol\twocolumn \else \newpage \fi
        \def\thefootnote{\arabic{footnote}}
        \setcounter{footnote}{0}}  

\catcode`@=12
\relax

\def\figcap{\section*{Figure Captions\markboth
        {FIGURECAPTIONS}{FIGURECAPTIONS}}\list
        {Figure \arabic{enumi}:\hfill}{\settowidth\labelwidth{Figure
999:}
        \leftmargin\labelwidth
        \advance\leftmargin\labelsep\usecounter{enumi}}}
 \relax
\def\tablecap{\section*{Table Captions\markboth
        {TABLECAPTIONS}{TABLECAPTIONS}}\list
        {Table \arabic{enumi}:\hfill}{\settowidth\labelwidth{Table
999:}
        \leftmargin\labelwidth
        \advance\leftmargin\labelsep\usecounter{enumi}}}
 \relax
\def\reflist{\section*{References\markboth
        {REFLIST}{REFLIST}}\list
        {[\arabic{enumi}]\hfill}{\settowidth\labelwidth{[999]}
        \leftmargin\labelwidth
        \advance\leftmargin\labelsep\usecounter{enumi}}}
 \relax
\makeatletter
\newcounter{pubctr}
\def\publist{\@ifnextchar[{\@publist}{\@@publist}}
\def\@publist[#1]{\list
        {[\arabic{pubctr}]\hfill}{\settowidth\labelwidth{[999]}
        \leftmargin\labelwidth
        \advance\leftmargin\labelsep
        \@nmbrlisttrue\def\@listctr{pubctr}
        \setcounter{pubctr}{#1}\addtocounter{pubctr}{-1}}}
\def\@@publist{\list
        {[\arabic{pubctr}]\hfill}{\settowidth\labelwidth{[999]}
        \leftmargin\labelwidth
        \advance\leftmargin\labelsep
        \@nmbrlisttrue\def\@listctr{pubctr}}}
 \relax
\makeatother
\newskip\humongous \humongous=0pt plus 1000pt minus 1000pt

\newif\ifdtup

\relax

\def\be{\begin{equation}}
\def\ee{\end{equation}}
\def\ba{\begin{eqnarray}}
\def\ea{\end{eqnarray}}

\def\a{\alpha}

\def\d{\delta}

\def\d{\textnormal{d}}

\def\no{\noindent}

\def\IR{\relax{\rm I\kern-.18em R}}
\def\II{\relax{\rm 1\kern-.35em1}}

\renewcommand{\theequation}{\thesection.\arabic{equation}}
\csname @addtoreset\endcsname{equation}{section}

\def\IR{\relax{\rm I\kern-.18em R}}
\def\inv{^{\raise.15ex\hbox{${\scriptscriptstyle -}$}\kern-.05em 1}}

\begin{document}

\begin{titlepage}
\begin{center}

\vskip .5in

{\LARGE One-loop quantization of rigid spinning strings in $AdS_3 \times S^3 \times T^4$ with mixed flux}
\vskip 0.4in

{\bf Juan Miguel Nieto} {and}
{\bf Roberto Ruiz}
\vskip 0.1in

Departamento de F\'{\i}sica Te\'orica \\
Universidad Complutense de Madrid \\
$28040$ Madrid, Spain \\
{\footnotesize{\tt juanieto@ucm.es, roruiz@ucm.es}}

\end{center}

\vskip .4in

\centerline{\bf Abstract}
\vskip .1in
\no

We compute the one-loop correction to the classical dispersion relation of rigid closed spinning strings with two equal angular momenta in the $AdS_3 \times S^3 \times T^4$ background supported with a mixture of R-R and NS-NS three-form fluxes. This analysis is extended to the case of two arbitrary angular momenta in the pure NS-NS limit. We perform this computation by means of two different methods. The first method relies on the Euler-Lagrange equations for the quadratic fluctuations around the classical solution, while the second one exploits the underlying integrability of the problem through the finite-gap equations. We find that the one-loop correction vanishes in the pure NS-NS limit.

\noindent

\vskip .4in
\noindent

\end{titlepage}

\vfill
\eject

\def\baselinestretch{1.2}


\baselineskip 20pt


\section{Introduction}

During recent years a renewed attention has been paid to the AdS$_3/$CFT$_2$ correspondence \cite{I,II} in $AdS_3 \times S^3\times {\cal M}_4$ backgrounds. The interest has been brought forth by the progress made in the application of integrability techniques developed for the AdS$_5$/CFT$_4$ correspondence \cite{review}. Its two best understood realizations involve string theories formulated on $AdS_3 \times S^3\times T^4$ \cite{VIII,SST,AB,SW,Ads3S3T4A,Borsato:2013hoa,Ads3S3T4B,BianchiHoare,Spinconnection,%
BothM4,completeworldsheet,Ads3S3T4C,BMNmismatch,Ads3S3T4D,BPS} and $AdS_3 \times S^3 \times S^3 \times S^1$ \cite{AB,SU11,SS,Rughoonauth,BOS,Abbott,BianchiHoare,SW,BothM4,EGGL} spaces, which have proven to be classically integrable and conformally invariant \cite{BSZ,wulff}. On the one hand, the dual CFT$_{2}$ to the former is believed to reduce to the Sym$^N(T^4)$ orbifold at certain point of its moduli space\cite{LarsenMartinec}, albeit deeper insight remains to be gained. On the other hand, the dual CFT$_{2}$ to the latter is still not clear and there exist different proposals \cite{XIV,Tong,EberhardtGopakumar}. Besides, integrability has been shown to be more fruitful in the $\mathcal{M}_4=T^4$ than in the $\mathcal{M}_4=S^3 \times S^1$ case, where it has been shown to emerge also in its CFT$_{2}$ side \cite{IntegrabilityintheCFT}.

In spite of its similarities with the $AdS_5\times S^5$ scenario, these two backgrounds exhibit some features that obstruct a straightforward integrability approach. The presence of massless excitations play a prominent role among them, as it seems to be responsible for the mismatch between the computation of the dressing factor for massive excitations performed via perturbative world-sheet calculations \cite{Supressedchiralities,Beccariafinitegap} and crossing equations \cite{Borsato:2013hoa,Ads3S3T4C}. This link was proposed for the $AdS_3 \times S^3 \times T^4$ space in \cite{Aniceto}, where it is argued that the lack of suppression of wrapping corrections involving massless modes associated to the $T^4$ factor could explain the discrepancy.

Unlike $AdS_5 \times S^5$, just supported by a Ramond-Ramond (R-R) five-form flux, both $AdS_3 \times S^3 \times T^4$ and $AdS_3 \times S^3 \times S^3 \times S^1$ can be deformed through the addition of a Neveu-Schwarz-Neveu-Schwarz (NS-NS) three-form flux, which mixes with the R-R one. Integrability, conformal invariance and kappa symmetry are not spoiled by the introduction of the NS-NS flux \cite{CagnazzoZarembo}. The mixed flux set-up has been also extensively studied in the literature \cite{HT,B.Hoare,BianchiHoare,Lloyd,SW2,M.Baggio,Modulispace}.

The aim of this paper is to compute the one-loop correction to the dispersion relation of rigid spinning string on $AdS_3 \times S^3\times T^4$ in the presence of mixed flux. To this end, we use two different methods. The first one extracts the characteristic frequencies from the Lagrangian of quadratic fluctuations, whose signed sum leads to the one-loop correction. This procedure has been already applied successfully to different string configurations in $AdS_5 \times S^5$ \cite{FirstQuantizing,Quantizing,Ghostcontribution,Quantizing2,Schafer,Nameki,SchaferNameki,dunne,Forini}. It was also applied to rigid spinning strings on $AdS_3 \times S^3\times T^4$ with pure R-R flux \cite{Supressedchiralities}. The second method starts from the construction of the algebraic curve associated to the Lax connection of the $PSU(1,1|2)^2/{(SU(1,1) \times SU(2))}$ supercoset sigma model. The quantization of the algebraic curve is well understood for the $AdS_5 \times S^5$ case (see \cite{FinitegapReview} for an in-depth explanation). The extension of this procedure to the $AdS_3 \times S^3\times T^4$ space is straightforward for the massive excitations with pure R-R flux. However, the construction of the finite-gap equations requires some extensions when dealing with the full spectrum supported by a general mixed flux. First of all, the introduction of an NS-NS flux shifts the poles of the Lax connection away from $\pm 1$. This issue has been solved in \cite{FinitegapBabichenko}, where the finite-gap equations proposed for $AdS_3 \times S^3\times T^4$ in the pure R-R regime \cite{BSZ} are generalized. A second problem arises from massless excitations, which require to loosen the usual implementation of the Virasoro constraints \cite{masslessfinitegap}. The one-loop correction to the BMN string for $AdS_3 \times S^3\times T^4$ and $AdS_3 \times S^3 \times S^3 \times S^1$ with pure R-R flux has been obtained within this framework in \cite{masslessfinitegap}, while the respective correction to the short folded string in such backgrounds has been derived in \cite{Beccariafinitegap}. The one-loop correction to the BMN string for $AdS_3 \times S^3\times T^4$ with mixed flux has been computed in \cite{FinitegapBabichenko}.

The outline is as follows. In section~\ref{classicalsetting} we summarize some features of the rigid classical spinning string solution on $\mathbb{R}\times S^3 \subset AdS_3 \times S^3 \times T^4$. The pure NS-NS limit and the restriction to the $\mathfrak{su}(2)$ sector are then discussed. In section~\ref{quadraticfluctuations} we compute the Lagrangian of the quadratic fluctuations around the rigid spinning solution and solve its equations of motion, hence obtaining their characteristic frequencies. In section~\ref{algebraiccurve} we check the results of the previous section by rederiving the characteristic frequencies from the flux-deformed finite-gap equations. In doing so, we neglect contributions from the massless excitations as we will have proven in section~\ref{quadraticfluctuations} that their net contribution ultimately vanishes. In section~\ref{comparison} we make use of the characteristic frequencies previously obtained to compute the one-loop correction to the dispersion relation. In section~\ref{non-rigid} we present an argument regarding the extension to non-rigid strings in the NS-NS limit. We close the article with summary and conclusions. In the appendices we have collected conventions and detailed computations that supplement the main text.

\section{The $\mathbb{R}\times S^3$ string with mixed flux}
\label{classicalsetting}

In this section we review the dynamics of rigid closed spinning strings on a $\mathbb{R}\times S^3 \subset AdS_3 \times S^3\times T^4$ background in the presence of both R-R and NS-NS three-form fluxes. This kind of solution as been studied, for example, in references \cite{Rotating1,HN1,HN2,Rotating2}. \footnote{The general spinning string solution encompasses other well known solutions as particular limits which are interesting enough to be studied separately. Examples are the giant magnon \cite{both,C.Ahn}, spiky strings \cite{both} and multi-spike strings \cite{multispike}. Spinning D1-strings have also been studied \cite{D1}.} After doing so, we restrict ourselves to two regimes of interest, the $\mathfrak{su}(2)$ sector and the pure NS-NS limit, which are the focus of the reminder of the article.

The $AdS_3\times S^3 \times T^4$ background metric can be parameterized by
\be
ds^2 = -dz_0^2 - z_0^2 \, dt^2 +dz_1^2 + z_1^2 \, d \phi^2 + 
dr_1^2 + r_1^2 d\varphi_1^2 + dr_2^2 +r_2^2 d \varphi_2^2 +\sum_{i=1}^4 d\gamma_i^2 \ ,
\ee
with the constraints
\begin{equation}
r_1^2 + r_2^2 = 1 \ , \quad \quad -z_0^2 + z_1^2 = -1 \ . 
\label{threesphere}
\end{equation}
Even though the background is supported by R-R and NS-NS fluxes, only the latter contribute to the bosonic string Lagrangian as the B-field
\be
\label{Kalb-Ramond}
B=q z_1^2 \, dt\wedge d\phi-q r_2^2 \, d\varphi_{1}\wedge d\varphi_2 \ ,
\ee
where the parameter $q\in[0,1]$ measures the mixing of the two fluxes.

In the case of pure R-R flux, $q=0$, the theory can be formulated in terms of a Metsaev-Tseylin action akin to that of the $AdS_5 \times S^5$ background \cite{Supercoset}. The Lagrangian associated to the bosonic spinning string ansatz in $AdS_3 \times S^3$ reduces in this case to the Neumann-Rosochatius integrable system \cite{NR}. Turning on the NS-NS flux introduces an additional term in the latter which does not spoil integrability \cite{HN1,HN2}. In fact, the complete Lagrangian remains integrable under this deformation \cite{CagnazzoZarembo}. The pure NS-NS flux limit, $q=1$, is of particular interest because the action can be re-expressed as a supersymmetric WZW model \cite{Moo} (see also the recent paper \cite{A.Dei} and references therein), which leads to several simplifications.

Instead of considering the most general setup, the remainder of this paper is devoted to bosonic classical spinning strings rotating in $S^3$ at the center of $AdS_3$ with no dynamics along $T^4$. Thus we fix $z_0=1$ and $\gamma_{i}=0$, and impose the ansatz describing a closed string rotating with two angular momenta in $S^3$  
\begin{align}
\label{ansatz}
t (\tau,\sigma) &= w_0 \tau \ , & r_i (\tau,\sigma) &= r_i (\sigma) \ , & \varphi_i (\tau,\sigma) &= \omega_i \tau + \alpha_i(\sigma) \ , &  i=1,2 \ .
\end{align}
Dealing with closed string solutions requires periodic boundary conditions of (\ref{ansatz}) on $\sigma$, entailing
\begin{equation}
	r_i (\sigma + 2 \pi) = r_i (\sigma) \ , \quad \quad  \alpha_i(\sigma + 2 \pi) = \alpha_i(\sigma) + 2 \pi {m}_i \ ,
\label{periodicity}
\end{equation}
where $m_i$ are integer winding numbers.
After substituting the ansatz (\ref{ansatz}) into the Polyakov action with the B-field in the conformal gauge, we obtain
\be
L_{S^3} = h \left[  \sum_{i=1}^2  \left( r_i'^2 + r_i^2 \a_i'^2 - r_i^2 \omega_i^2 \right) 
- {\Lambda} \left( \sum_{i=1}^2r_i^2 - 1\right) + 2q r_2^2 \, \left( \omega_1 \alpha_2' - \omega_2 \alpha_1' \right) \right] \ ,
\label{NRq}
\ee
where the prime stands for derivatives with respect to $\sigma$ and $h=h(\sqrt{\lambda})$ is the coupling constant. \footnote{The relationship between the coupling constant and the string tension is the same as the one for $AdS_5 \times S^5$ at first order, i.e. $h= \frac {\sqrt{\lambda}}{4 \pi}+\dots$, although it might receive corrections (both perturbative and non-perturbative) in the 't Hooft coupling. However, it is known that the first correction $\mathcal{O} (1)$ should vanish for the pure RR case \cite{wulff,Abbott,Supressedchiralities} and it might vanish also for the mixed flux case.} In addition, we can construct three non-vanishing conserved charges from the isometries of the metric: the energy 
\begin{align}
E & = 4\pi h w_0 \ , \label{E}
\end{align}
and the two angular momenta
\begin{align}
J_1 & = 2h\int_0^{2\pi} {d \sigma \left( r^2_1 \omega_1 - q r_2^2 \alpha_2' \right)} \ , \label{J1} \\
J_2 & = 2h\int_0^{2 \pi} {d \sigma \left( r^2_2 \omega_2 + q r_2^2 \alpha_1' \right)} \ . \label{J2}
\end{align}

The Euler-Lagrange equations for the radial coordinates are
\begin{align}
r_1'' & = -r_1 \omega_1^2 +r_1 \alpha_1^{\prime 2}-\Lambda r_1 \ , \label{r1prime} \\
r_2'' & = -r_2 \omega_2^2 +r_2 \alpha_2^{\prime 2}-\Lambda r_2 + 2 q r_2 ( \omega_1 \alpha_2' - \omega_2 \alpha_1' ) \ . \label{r2prime}
\end{align}
On the other hand, the equations of motion for the angular equations can be easily integrated as the Lagrangian only depends on their derivatives
\begin{equation}
\alpha_1' = \frac {v_1 + q r_2^2 \omega_2}{r_1^2} \ , \quad \quad \alpha_2' = \frac {v_2 - q r_2^2 \omega_1}{r_2^2} \ ,
\label{alphaprime}
\end{equation}
where $v_i$ are integration constants which can be understood as the momenta associated to $\alpha_i$. Replacing $\alpha '_i$ by these momenta in the Lagrangian (\ref{NRq}) leads to aforementioned deformation of the Neumann-Rosochatius system. The equations of motion must be supplemented with the Virasoro constraints, which read
\begin{align}
& \sum_{i=1}^2 \left[ r_i^{\prime 2} + r_i^2 ( \alpha^{\prime 2}_i + \omega_i^2 ) \right] = w_0^2 \ , \label{Virasoro1} \\
& \sum_{i=1}^2 r_i^2 \omega_i \alpha_i'= \sum_{i=1}^2 v_i \omega_i =0 \ . \label{Virasoro2} 
\end{align}

The most straightforward solutions to the equations of motion are those of constant radii
\begin{equation}
	r_i (\sigma )=a_i \label{Constant radii} \ .
\end{equation}
On these solutions $\alpha '_i=m_i$ and (\ref{r1prime}) and (\ref{r2prime}) simplify to
\begin{equation}
\label{Lagrange multiplier}
\Lambda=m_{1}^2-\omega_{1}^2=m_{2}^2-\omega_{2}^2+2q(\omega_1 m_2-\omega_2 m_1) \ .
\end{equation}
Furthermore, the restriction (\ref{threesphere}) and the Virasoro constraint (\ref{Virasoro2}) determine
\begin{equation}
\label{Radii}
a_1^2 =\frac{-\omega_2 m_2}{\omega_1 m_1-\omega_2 m_2} \ , \quad \quad a_2^2 =\frac{\omega_1 m_1}{\omega_1 m_1-\omega_2 m_2} \ ,
\end{equation}
or, using the definitions of the angular momenta,
\begin{equation}
a_1^2 =\frac{J_1 +4\pi hq m_2}{4\pi h (\omega_1 +qm_2)} \ , \quad \quad a_2^2 =\frac{J_2}{4\pi h (\omega_2 +qm_1)} \label{Angular momenta} \ .    
\end{equation}

The dispersion relation for general values of $q$ only can be written down as a series in negative powers of the total angular momentum. Despite so, there exist two regimes where we can find a closed analytic expression for the dispersion relation. The first one is the pure NS-NS limit we commented above, while the second one corresponds to a restriction to a $\mathfrak{su}(2)$ subsector of the theory, which amounts to set the $J_1=J_2$. Note that these two regimes are not mutually exclusive.

Let us examine in more depth both cases. In the first limit the equations of motion can be solved by
\begin{equation}
	\omega_1 =\frac{J}{4\pi h} \ , \quad \quad  \omega_2 =\frac{J}{4\pi h}-(m_1 - m_2) \ ,
\end{equation}
where $J=J_1+J_2$. Using the first Virasoro constraint (\ref{Virasoro1}) and (\ref{Radii}) we can obtain the dispersion relation
\begin{equation}
\label{Dispersion relation}
	E=J+4\pi h m_1 \ .
\end{equation}

On the other hand, if we restrict ourselves to the $\mathfrak{su}(2)$ subsector, the second Virasoro constraint (\ref{Virasoro2}) implies $m_{1}=-m_{2}\equiv m$, while the equations of motion and the definition of the angular momenta imposes
\begin{equation}
\label{Primitive frequencies}
\omega_1 =\frac{J}{4\pi h}=\Upsilon+qm \ , \quad \quad \omega_2 =\frac{J}{4\pi h}-2qm=\Upsilon-qm \ ,
\end{equation}
where we have defined
\begin{equation}
	\Upsilon=\frac{J}{4\pi h}-qm=\frac{J_1+J_2}{4\pi h}-qm \ ,
\end{equation}
for later convenience. The expression for the dispersion relation reads
\begin{equation}
	E=\sqrt{J^2-8\pi h q m J +16\pi^2 h^2 m^2 }=4\pi h \sqrt{\Upsilon^ 2 + \kappa^2 m^2 } \ ,
\end{equation}
with $\kappa^2=1-q^2$.

\section{Quadratic fluctuations around the classical solution}
\label{quadraticfluctuations}

In this section we derive the Lagragian for the quadratic fluctuations around the rigid spinning string-type solutions on $\mathbb{R}\times S^3$. We obtain the equation for the characteristic frequencies for them, which greatly simplifies in both the case of two equal angular momenta and the case of pure NS-NS flux. The signed sum of the characteristic frequencies provide the one-loop correction to the classical energy.

We derive this effective Lagrangian by splitting the target-space fields into the classical background and fluctuation fields and truncating the latter up to second order in the action. We treat the bosonic fluctuations on the sphere, on the anti-de Sitter space, on the torus and the fermionic fluctuations separately in order to make the section more \break 
readable. \footnote{In this section we do not label the origin of the frequencies ($S$, $AdS$, $T$ or $F$) as no ambiguity arises.}

\subsection{Bosonic fluctuations on $S^ 3$}

Regarding the fluctuations on the sphere, we take advantage of the spherical symmetry and perform the following substitution
\begin{align}
\label{Fluctuating fields}
r_i \cos \varphi_i &\rightarrow a_i \cos (\alpha_i +\omega_i \tau) + \tilde{r}_i \cos (\alpha_i +\omega_i \tau) -\rho_i \sin (\alpha_i +\omega_i \tau) \ , \notag\\
r_i \sin \varphi_i &\rightarrow a_i \sin (\alpha_i +\omega_i \tau)+\tilde{r}_i \sin (\alpha_i +\omega_i \tau) +\rho_i \cos (\alpha_i +\omega_i \tau) \ .
\end{align}
where $\tilde{r_i},\rho_i$ denote the perturbation fields. \footnote{It is also possible to incorporate the fluctuations as $r_i \rightarrow a_i+\tilde{r}_i$ and $\varphi_i \rightarrow \alpha_i+\omega_i \tau +\tilde{\varphi}_i$ instead. Both choices are equivalent since $\rho_i \approx a_i \tilde{\varphi}_i$ at first order.}

Introducing (\ref{Fluctuating fields}) into the Polyakov action with the B-field in the conformal gauge, the Lagrangian for the quadratic fluctuations of the fields turns out to be

\begin{align}
\label{Lagrangian density}
	\tilde{L}_{S^3}&=h \left\{ \sum_{i=1}^2 \Big[ -\dot{\tilde{r}}_i^2 -\dot{\rho}_i^2+\tilde{r}_i^{\prime 2}+\rho_i^{\prime 2} +(\tilde{r}_i^2 +\rho_i^2)(-\omega_i^2+\alpha_i^{\prime 2} ) +2 \tilde{r}_i ( -\omega_i \dot{\rho}_i+\alpha_i^{\prime} \rho '_i) \right. \notag \\
	&-2\rho_i (-\omega_i \dot{\tilde{r}}_i+\alpha '_i \tilde{r}'_i) -\Lambda (\tilde{r}_i^2 + \rho_i^2)  \Big] +2q \bigg[ (\tilde{r}_2^2 +\rho_2^2) (\omega_1 \alpha '_2 -\alpha '_1 \omega_2)  \notag \\
	&+\tilde{r}_2 (\omega_1 \rho '_2 -\alpha '_2 \dot{\rho}_2 ) -\rho_2 (\omega_1 \tilde{r}'_2 -\alpha '_1 \dot{\tilde{r}}_2 )+\frac{a_2}{a_1} \left( 2\tilde{r_2} -\frac{a_2}{a_1} \tilde{r}_1 \right) (\dot{\rho}_1 \alpha '_2 -\rho '_1 \omega_2)  \notag \\
	&\left. -\frac{a_2^2}{a_1^2} \rho_1 (\dot{\tilde{r}}_1 \alpha '_2 -\tilde{r}'_1 \omega_2) +\frac{a_2}{a_1} (\dot{\rho}_1 \rho '_2 -\rho '_1 \dot{\rho}_2 ) \right]\Bigg\} \ ,
\end{align}
where the dot stands for derivatives with respect to $\tau$.

Besides, an orthogonality requirement has to be satisfied between the classical solution and the fluctuation fields. Such condition can be seen as a consequence of the perturbation of the Lagrange multiplier in (\ref{NRq}), promoted as $\Lambda\rightarrow\Lambda+\tilde{\Lambda}$. In particular, this constraint reads 
\begin{equation}
\label{Orthogonality}
\sum_{i=1}^2{a_{i}\tilde{r}_i}=0 \ .    
\end{equation}
Imposing it, the Euler-Lagrange equations for $\tilde{r}_2$, $\rho_1$ and $\rho_2$ become
\begin{align}
\label{PDE system}
	&-\ddot{\rho}_1+\rho ''_1+2\frac{a_2}{a_1} \left[ (\omega_1 +qm_2) \dot{\tilde{r}}_2- (m_1 +q\omega_2) \tilde{r}'_2 \right]=0 \ , \notag \\
	&-\ddot{\rho}_2+\rho ''_2 -2\left[ (\omega_2 +qm_1) \dot{\tilde{r}}_2- (m_2 +q\omega_1 ) \tilde{r}'_2 \right]=0 \ , \notag \\
	&-\frac{\ddot{\tilde{r}}_2}{a_{1}^2}+\frac{{\tilde{r}''}_2}{a_{1}^2}-2\frac{a_2}{a_1} \left[ (\omega_1 +q m_2) \dot{\rho}_1 - (m_1 +q\omega_2 ) \rho '_1 \right]\notag\\
	&+2 \left[ (\omega_2 +q m_1) \dot{\rho}_2 - (m_2 +q\omega_1 ) \rho '_2 \right]=0 \ .
\end{align}
We are allowed to decompose $\tilde{r}_{2}$, $\rho_1$ and $\rho_2$ in a base of exponential functions due to the periodic boundary conditions in $\sigma$. Supplementing the decomposition with an expansion in Fourier modes in $\tau$, we write
\begin{align}
\label{Expansion}
	\rho_j &=\sum_{n=-\infty}^{\infty}\sum_{k=1}^6A_{j,n}^{(k)}e^{i\omega_{k,n}\tau+in\sigma} \ , &
	\tilde{r}_2&=\sum_{n=-\infty}^{\infty}\sum_{k=1}^6B_{n}^{(k)}e^{i\omega_{k,n}\tau+in\sigma} \ .
\end{align}
The sum over $k$ follows from the existence of six different frequencies associated to the same mode number $n$ \cite{Characteristicfrequencies}. Employing this ansatz, the Euler-Lagrange equations (\ref{PDE system}) become the matrix equation
\begin{equation}
\label{Firstmatrix}
\begin{pmatrix}
\omega_{k,n}^2-n^2 & 0 & M_{13}  \\
0 & \omega_{k,n}^2-n^2 & M_{23}\\  
M_{31} & M_{32}  & \omega_{k,n}^2-n^2 \\
\end{pmatrix}
\begin{pmatrix}
A_{1,k,n}\\
A_{2,k,n}\\
B_{k,n}
\end{pmatrix}
\equiv M
\begin{pmatrix}
A_{1,k,n}\\
A_{2,k,n}\\
B_{k,n}
\end{pmatrix}
=0 \ .
\end{equation}
where the unspecified matrix elements are explicitly
\begin{align}
&M_{13}=2i \left[ (\omega_1 +qm_2) \omega_{k,n}- (m_1 +q\omega_2) n \right]\frac{a_{2}}{a_{1}} \ ,\notag \\
&M_{23}=-2i\left[ (\omega_2 +qm_1) \omega_{k,n}- (m_2 +q\omega_1 ) n\right] \ ,\notag \\
&M_{31}=-2i \left[ (\omega_1 +qm_2) \omega_{k,n}- (m_1 +q\omega_2) n \right]a_{1}a_{2} \ ,\notag \\
&M_{32}=2i\left[ (\omega_2 +qm_1) \omega_{k,n}- (m_2 +q\omega_1 ) n\right]a_{1}^2 \ .
\end{align}
The existence of non-trivial solutions to (\ref{Firstmatrix}) requires $\text{det}\, M=0$, which provides the following characteristic equation for the frequencies
\begin{align}
\label{Characteristic equations}
&(\omega_{k,n}^2-n^2) \left\{(\omega_{k,n}^2-n^2)^2-4 a_{1}^2 \ [ (\omega_1+qm_2) \omega_{j,n} -(m_1+q\omega_2)n]^2 \right. \notag \\
&\left.-4a_{2}^2 \ [(\omega_2+q m_1)\omega_{k,n}-(m_2+q\omega_1)n]^2\right\}=0 \ .
\end{align}
This equation has six solution as we commented above, and reduces to the one obtained in \cite{Quantizing} in the limit of a pure R-R flux.

We should note that two frequencies corresponding to decoupled massless modes arise as solutions to the characteristic equation. They emerge as a consequence of the conformal gauge-fixing condition \cite{Ghostcontribution}. Solving the linearized Virasoro constraints for the $AdS_ {3}$ decoupled massless field and substituting the solution back in the full Lagrangian remove these spurious frequencies from (\ref{Characteristic equations}), since their associated field gets cancelled. Therefore, we can safely ignore them.



In principle, we can find the remaining solutions to the equation (\ref{Characteristic equations}) as a series in inverse powers of the total angular momentum $J$. Since expressions thus obtained are not very enlightening, we focus exclusively on the two regimes we presented at the end of the previous section.

In the $\mathfrak{su}(2)$ sector ($m_1=-m_2=m$) the frequencies, written as a series in $\Upsilon$, are
\begin{align}
	\omega_{1,n} &= 2\Upsilon -nq +\frac{n^2 \kappa^2}{2\Upsilon}+\frac{q \kappa^2 n(n^2-2m^2)}{2\Upsilon^2} +\mathcal{O} \left( \frac{1}{\Upsilon^3} \right) \ , \notag \\
	\omega_{2,n} &= -2\Upsilon -nq -\frac{n^2 \kappa^2}{2\Upsilon}+\frac{q \kappa^2 n(n^2-2m^2)}{2\Upsilon^2} +\mathcal{O} \left( \frac{1}{\Upsilon^3} \right)\ , \notag \\
	\omega_{3,n} &= nq + \frac{n \kappa^2 \sqrt{n^2 -4m^2}}{2\Upsilon}-\frac{q \kappa^2 n(n^2-2m^2)}{2\Upsilon^2} +\mathcal{O} \left( \frac{1}{\Upsilon^3} \right) \ , \notag \\
	\omega_{4,n} &= nq - \frac{n \kappa^2 \sqrt{n^2 -4m^2}}{2\Upsilon}-\frac{q \kappa^2 n(n^2-2m^2)}{2\Upsilon^2} +\mathcal{O} \left( \frac{1}{\Upsilon^3} \right) \label{quadraticfrequenciesS}\ ,
\end{align}
whereas in the pure NS-NS limit they read 
\begin{alignat}{2}
	\omega_{1,n} &= 2\left( \frac{J}{4\pi h} +m_2 \right) -n \ , &\quad \quad \omega_{3,n} &= n \ , \notag \\
	\omega_{2,n} &= -2\left( \frac{J}{4\pi h} +m_2 \right) -n \ , &\quad \quad \omega_{4,n} &= n \ .
\end{alignat}
In the overlap of both regimes the frequencies can be written as follows
\begin{alignat}{2}
\label{Fre}
	\omega_{1,n} &=  2\left(\frac{J}{4\pi h}-m\right)-n= 2\Upsilon-n \ , &\quad \quad \omega_{3,n}&=n \ , \\
	\omega_{2,n} &= -2\left(\frac{J}{4\pi h}-m\right)-n=-2\Upsilon-n \ , &\quad \quad \omega_{4,n}&=n \ . \notag
\end{alignat}

\subsection{Bosonic fluctuations on $AdS_3$}

We proceed analogously for $AdS_3$ fluctuations using the parameterization
\begin{align}
	z_0 \cos t &\rightarrow (1+\tilde{z}_{0})\cos (\kappa \tau)-\chi_{0}\sin (\kappa \tau) \ , & z_1 \sin \phi &\rightarrow \tilde{z}_{1} \ , \notag \\
	z_0 \sin t &\rightarrow(1+\tilde{z}_{0})\sin (\kappa \tau)+\chi_{0}\cos (\kappa \tau) \ , & z_1 \cos \phi &\rightarrow \chi_{1} \ .
\end{align}
This choice leads to the following Lagrangian density for the fluctuation fields
\begin{align}
\label{Anti-de Sitter Lagrangian density}
	\tilde{L}_{AdS_3} &=h \left[ \tilde{z}_{0}'^2+\chi_{0}'^2-\dot{\tilde{z}}_{0}^2-\dot{\chi}_{0}^2
	-\tilde{z}_{1}'^2-\chi_{1}'^2 +\dot{\tilde{z}}_{1}^2 +\dot{\chi}_{1}^2 +2w_0 \left( \tilde{z_{0}}\dot{\chi}_{0}-\chi_0\dot{\tilde{z}}_{0} \right) \right. \notag \\
	& \left. +w_0^2(\tilde{z}_{1}^2+\chi_{1}^2) -2qw_0 (\tilde{z}_{1} \chi'_1-\chi_1 \tilde{z}'_1) \right]  \ .
\end{align}
This Lagrangian has to be supplemented with an orthogonality constraint between the background and the fluctuation fields similar to (\ref{Orthogonality}), which in this case reads $\tilde{z}_0=0$. As a consequence, the field $\chi_0$ decouples, leading to two massless excitations that can be ignored in view of the discussion above. The relevant Euler-Lagrange equations are
\begin{align}
-\ddot{\tilde{z}}_{1}+{\tilde{z} ''_{1}}-w_0^2\tilde{z}_{1}+2 q w_0\chi_{1}' &=0 \ , &
-\ddot{\chi}_{1}+{\chi ''_{1}}-w_0^2\chi_{1}-2qw_0\tilde{z_{1}}' &=0 \ .
\end{align}
An analogous expansion to (\ref{Expansion}) allows us to derive the characteristic equations for the frequencies. In this case, 
\begin{equation}
(\omega_{k,n}^2-n^2-w_0^2)^2-4q^2n^2w_0^2=0 \ ,    
\end{equation}
whose solutions are
\begin{align}
	\omega_{1,n} &=\sqrt{n^2+2qw_0 n+w_0^2} \ , & \omega_{2,n} &=\sqrt{n^2-2qw_0 n+w_0^2} \ ,\notag \\
	\omega_{3,n} &=-\sqrt{n^2+2qw_0 n+w_0^2} \ , & \omega_{4,n} &=-\sqrt{n^2-2qw_0 n+w_0^2} \ . \label{quadracticfrequenciesads} \ .    
\end{align}
All of them reduce to already known result in the non-deformed case \cite{Supressedchiralities,Quantizing}. Furthermore, the pure NS-NS limit allow us to complete squares, obtaining
\begin{equation}
\label{cuen}
\omega_{k,n}=\pm n \pm w_0 \ ,
\end{equation}
with uncorrelated signs.

\subsection{Bosonic fluctuations on $T^4$}

Since we consider no classical dynamics on the torus, we are led to a free Lagrangian for the fluctuations. Therefore, the characteristic frequencies are \footnote{In the $T^4$ space one could also consider fluctuations with non-trivial windings. However, after averaging over these windings, one is left only with the contribution from the zero winding sectors. We want to thank Tristan McLoughlin for pointing us this issue.}
\begin{equation}
\omega_{k,n}= \pm n \ ,
\end{equation}
with one pair of solutions for each of the four coordinates.

\subsection{Fermionic fluctuations}

As our background solution is purely bosonic, the Lagrangian for the fermionic fluctuations reduces to the usual fermionic Lagrangian computed up to quadratic order in the fermionic fields. For a type IIB theory, the latter is given by
\begin{equation}
\label{Fermionic Lagrangian density}
\tilde{L}_{F}=i(\eta^{\alpha\beta}\delta_{\dot{I}\dot{J}}-\epsilon^{\alpha\beta}(\sigma_{3})_{\dot{I}\dot{J}})\bar{\theta}^{\dot{I}}\rho_{\alpha}{\left(\textnormal{D}_{\beta}\right)^{\dot{J}}}_{\dot{K}}\theta^{\dot{K}} \ ,
\end{equation}
where the covariant derivative is \cite{Spinconnection,completeworldsheet}
\begin{equation}
\label{Covariant derivative}
{(\textnormal{D}_{\alpha})^{\dot{I}}}_{\dot{J}}={\delta^{\dot{I}}}_{\dot{J}}\left(\partial_{\alpha}-\frac{1}{4}\omega_{\alpha ab}\Gamma^{a}\Gamma^{b}\right)+\frac{1}{8}{\left(\sigma_{1}\right)^{\dot{I}}}_{\dot{J}}e^{a}_{\alpha}H_{abc}\Gamma^{b}\Gamma^{c}+\frac{1}{48}{{\left(\sigma_{3}\right)^{\dot{I}}}_{\dot{J}}}F_{abc}\Gamma^{a}\Gamma^{b}\Gamma^{c} \ .
\end{equation}
We refer to appendix~\ref{conventions} for definitions and conventions.

We get the characteristic equation for the frequencies by substituting the classical solution, fixing the kappa symmetry and expanding the fermions in Fourier modes. In particular, the kappa gauge condition we choose is  $\theta^{1}=\theta^{2}\equiv\theta$ as \cite{BSZ,Supressedchiralities}. The Fourier expansion reads
\begin{equation}
\label{Grassmannian expansion}
\theta=\overset{\infty}{\underset{n=-\infty}{\sum}} \overset{8}{\underset{k=1}{\sum}} \theta_{n}^{(k)}e^{i\omega_{k,n} \tau +in\sigma} \ , 
\end{equation}
where we have summed up to eight frequencies for each mode instead of up to the sixteen frequencies that would be expected from the number of degrees of freedom \cite{Fermionicdegreesoffreedom}. We are allowed to do so because only six of the ten target-space coordinates are non-trivially involved in the equations of motion and hence we can restrict ourselves to six-dimensional gamma matrices. To recover the full set of frequencies we have to double the multiplicity of each frequency $\omega_{k,n}$. We should remark that we have imposed periodic boundary conditions to the fermionic fields as \cite{Quantizing,BTK,HernandezLopez}, relying on the discussion of the appendix E of \cite{FinitegapGromov}. Imposing the vanishing of the determinant of the differential operator in (\ref{Fermionic Lagrangian density}), the resulting expression for the characteristic equation is
\begin{align}
	\frac{m^8}{256} &\left[ m^2 \kappa^2 (4\kappa^2 -5) +4n^2 +(3-4q^2) \Upsilon^2 +4 \omega_{k,n} (w_0-\omega_{k,n}) \right] \notag \\
	\times &\left[ m^2 \kappa^2 (4\kappa^2 -5) +4n^2 +(3-4q^2) \Upsilon^2 -4 \omega_{k,n} (w_0+\omega_{k,n}) \right] \notag \\
	\times &\left[ m^2 \kappa^2 (4\kappa^2 -3) -4n^2 +(1-4q^2) \Upsilon^2 +8qn w_0 +4 \omega_{k,n} (\omega_{k,n} -w_0) \right] \notag \\
	\times &\left[ m^2 \kappa^2 (4\kappa^2 -3) -4n^2 +(1-4q^2) \Upsilon^2 -8qn w_0 +4 \omega_{k,n} (\omega_{k,n} +w_0) \right]=0 \ ,
\end{align}
where $\kappa^2=1-q^2$. Note that it is a polynomial of eighth degree, which agrees with the discussion above. 

Solving this equation, we find the following frequencies for the $\mathfrak{su}(2)$ sector:
\begin{align}
\label{Fermionic frequencies}
	\omega_{1,n} &=n+\left(q-\frac{1}{2}\right) w_0 \ , & \omega_{2,n} &=n-\left( q-\frac{1}{2}\right) w_0 \ , \notag \\
	\omega_{3,n} &=-n+\left(q+\frac{1}{2}\right) w_0 \ , & \omega_{4,n} &=-n-\left(q+\frac{1}{2}\right) w_0 \ , \notag \\
	\omega_{5,n} &=\sqrt{n^2-q^2 w_0^2+\Upsilon^2}+\frac{1}{2} w_0 \ , & \omega_{6,n} &=\sqrt{n^2-q^2 w_0^2+\Upsilon^2}-\frac{1}{2} w_0 \ , \notag \\
	\omega_{7,n} &=-\sqrt{n^2-q^2 w_0^2+\Upsilon^2}+\frac{1}{2} w_0 \ , & \omega_{8,n} &=-\sqrt{n^2-q^2 w_0^2+\Upsilon^2}-\frac{1}{2} w_0 \ .
\end{align}
We stress that the massless frequencies remain as such independently from the mixing parameter, cancelling the contributions from the $T^4$ modes for all values of $q$. Performing the $q\rightarrow 1$ limit manifestly simplifies the expressions in (\ref{Fermionic frequencies}) \footnote{In this limit we can find and solve the equation beyond the $\mathfrak{su}(2)$ sector. The difference amounts to the following shifts of the frequencies: $\omega_i=(m_1+m_2)+\omega^{\mathfrak{su}(2)}_i$ for $i=3,7$, $\omega_i=-(m_1+m_2)+\omega^{\mathfrak{su}(2)}_i$ for $i=4,8$ and $\omega_i=\omega^{\mathfrak{su}(2)}_i$ for the remainder.}
\begin{align}
\label{cias}
	\omega_{1,n} &=n+\frac{1}{2} w_0 \ , & \omega_{2,n} &=n-\frac{1}{2} w_0 \ ,& \omega_{3,n} &=-n+\frac{3}{2} w_0 \ , & \omega_{4,n} &=-n-\frac{3}{2} w_0 \ ,\nonumber \\
	\omega_{5,n} &=n+\frac{1}{2} w_0 \ , & \omega_{6,n} &=n-\frac{1}{2} w_0 \ , & \omega_{7,n} &=-n+\frac{1}{2} w_0 \ , & \omega_{8,n} &=-n-\frac{1}{2} w_0 \ ,
\end{align}
where we have used that $\lim_{q\rightarrow 1} w_0=\Upsilon$.

\section{Frequencies from the algebraic curve}
\label{algebraiccurve}

In order to check the computations performed in the previous section, we derive again the frequencies associated to the fluctuations using a different method that relies on the integrability of our problem: the semi-classical quantization of the classical algebraic curve. We start by constructing the eigenvalues of the monodromy matrix for the classical solution, whose associated quasi-momenta define a Riemann surface. This classical setting is quantized by adding infinitesimal cuts to this surface, thus modifying the analytical properties of the quasi-momenta, which contain the one-loop correction to the energy.

The Riemann surface we are interested in presents only one cut, analogously to the one studied for the $AdS_5 \times S^5$ scenario in \cite{FinitegapGromov}. Although the presence of the NS-NS flux deforms the construction of the algebraic curve, hence rendering the results from $AdS_5 \times S^5$ inapplicable, the procedure remains mostly the same and can be used as guideline. For the full detailed derivation of the finite-gap equations for general values of $q$ we refer to \cite{FinitegapBabichenko}, where the authors apply them to the particular case of the BMN string. Although this procedure neglects the contribution from the massless excitations, we already know from the previous section that their contributions cancel each other.

We start the computation of the classical algebraic curve by choosing the gauge $g_L \oplus g_R=g\oplus 1\in \mathfrak{psu} (1,1|2)^2$ where \footnote{Notice that the choice of representative in \cite{FinitegapBabichenko} was not correct. Even though $a_i^2=\frac{J_i}{w_i}$ holds for the undeformed case, this relation gets modified due to the flux becoming the one shown in equation (\ref{Angular momenta}). In any case, this problem does not affect the computation of the correction to the BMN spectrum therein. Our quasi-momenta for the rigid spinning string are the same as those obtained there after replacing their $\Omega$ by our $\Upsilon$.}

\begin{equation}
g=\left( \begin{matrix}
e^{i w_0 \tau} & 0 & 0 & 0 \\
0 & e^{-i w_0 \tau} & 0 & 0 \\
0 & 0 & a_1 e^{i (\omega_1 \tau +m_1 \sigma)} & a_2 e^{i (\omega_2 \tau +m_2 \sigma)} \\
0 & 0 & -a_2 e^{-i (\omega_2 \tau +m_2 \sigma)} & a_1 e^{-i (\omega_1 \tau +m_1 \sigma)}
\end{matrix}\right) \ .
\end{equation}
Using the normalization from \cite{FinitegapBabichenko} \footnote{We have chosen this normalization over the one proposed in \cite{CagnazzoZarembo} because it facilitates our later construction of the corrections to the classical curve by simplifying the relation between the Lax connection and the Noether currents.} the Lax connection associated reads
\begin{align}
	\mathcal{L} (x)&=\left( \begin{matrix}
		\hat{\cal L} (x) & 0 \\
		0 & \check{\cal L} (x)
	\end{matrix} \right)=\left( \begin{matrix}
		\hat{\cal L} (x) & 0 \\
		0 & \hat{\cal L} \left(\frac{1}{x} \right)
	\end{matrix} \right) \ , \\
	\hat{\mathcal{L}} (x)&=\frac{ix}{(x-s) \left( x+\frac{1}{s} \right)} \left( \begin{matrix}
w_0/\kappa & 0 & 0 & 0 \\
0 & -w_0/\kappa & 0 & 0 \\
0 & 0 & -m\left( \frac{q}{\kappa} -x \right) & e^{-2im (\sigma +q\tau)} \frac{\sqrt{\Upsilon^2 -m^2 q^2}}{\kappa} \\
0 & 0 & e^{2im (\sigma +q\tau)} \frac{\sqrt{\Upsilon^2 -m^2 q^2}}{\kappa} & m\left( \frac{q}{\kappa} -x \right)
\end{matrix}\right) \notag \ .
\end{align}
Notice that the usual $\pm 1$ poles of the Lax connection shift due to the presence of both fluxes, appearing now at $\pm s$ and $\pm \frac{1}{s}$, where $s=s(q)$ is defined as
\begin{equation}
	s=\sqrt{\frac{1+q}{1-q}} \ .
\end{equation}
The quasi-momenta associated to this Lax connection, obtained from the logarithm of the eigenvalues of the monodomy matrix, are
\begin{align}
	\hat{p}_1^A (x)&=-\hat{p}_2^A (x)=\check{p}_1^A \left( \frac{1}{x} \right)=-\check{p}_2^A \left( \frac{1}{x} \right)=\frac{2\pi x w_0}{\kappa (x-s) \left( x+\frac{1}{s} \right)} \label{adsquasimomenta} \ , \\
	\hat{p}_1^S (x)&=-\hat{p}_2^S (x)=\frac{2\pi x K(1/x)}{\kappa (x-s) \left( x+\frac{1}{s} \right)} \label{shatquasimomenta}  \ , \\
	\check{p}_1^S (x)&=-\check{p}_2^S (x)=-\frac{2\pi x K(x)}{\kappa (x+s) \left( x-\frac{1}{s} \right)}+2\pi m \label{scheckquasimomenta} \ ,
\end{align}
where $K(x)=\sqrt{m^2 x^2 \kappa^2+2q\kappa m^2 x +\Upsilon^2}$.
In order to find the one-loop correction to the energy we add extra cuts to the Riemann surface. These cuts are infinitesimally small and appear as poles on the quasi-momenta. Depending on the sheets of the Riemann surface the infinitesimal cuts connect, they correspond to different kinds of excitations. The precise relations between them are
\begin{align*}
	\text{AdS}_3 & : \quad (\hat{p}_1^A , \hat{p}_2^A),  \, (\check{p}_2^A , \check{p}_1^A) \ ,\\
	\text{S}^3 & : \quad  (\hat{p}_1^S , \hat{p}_2^S),  \, (\check{p}_2^S , \check{p}_1^S) \ ,\\
	\text{Fermions} & : \quad  (\hat{p}_1^A , \hat{p}_2^S),  \, (\hat{p}_1^S , \hat{p}_2^A),  \, (\check{p}_2^A , \check{p}_1^S),  \, (\check{p}_2^S , \check{p}_1^A)  \ .
\end{align*}
The explicit residues of these poles are
\begin{align}
	\res_{x=\hat{x}_n^{AX}} \delta \hat{p}_i^A &=-(\delta_{1i} -\delta_{2i}) \hat{\alpha} (\hat{x}_n^{AX}) N_n^{AX} \ , &
	\res_{x=\hat{x}_n^{SX}} \delta \hat{p}_i^S &=+(\delta_{1i} -\delta_{2i}) \hat{\alpha} (\hat{x}_n^{SX}) N_n^{SX} \ , \notag \\
	\res_{x=\check{x}_n^{AX}} \delta \check{p}_i^A &=+(\delta_{1i} -\delta_{2i}) \check{\alpha} (\check{x}_n^{AX}) N_n^{AX} \ , &
	\res_{x=\check{x}_n^{SX}} \delta \check{p}_i^S &=-(\delta_{1i} -\delta_{2i}) \check{\alpha} (\check{x}_n^{SX}) N_n^{SX} \label{checkpoles} \ ,
\end{align} 
where $X$ is either $A$ or $S$ depending on which sheet the cut ends. The functions $\hat{\alpha}(x)$ and $\check{\alpha}(x)$ are defined as
\begin{equation}
	\hat{\alpha} (x) =\frac{x^2}{\kappa h(x-s) \left( x+\frac{1}{s} \right)} \ , \quad \quad \check{\alpha} (x) =\frac{x^2}{\kappa h(x+s) \left( x-\frac{1}{s} \right)} \ .
\end{equation}
To fix the positions where we have to add these poles we have to use the relation between the quasi-momenta above and below the branch cuts of the Riemann surface $C_{ij}$, where $i$ and $j$ (comprising both $A$ or $S$ and $1$ or $2$) label the sheets the cut connects,
\begin{equation}
	p_i^+(x) -p_j^- (x)=2\pi n \ , \qquad x\in C_{ij}\ .
\end{equation}
This equation not only constraints the positions of the poles $x_n^{ij}$, but also the behaviour of the corrections to the quasi-momenta on them
\begin{equation}
	p_i (x_n^{ij}) - p_j (x_n^{ij}) =2\pi n \ , \quad \quad (\delta p_i)^+ (x_n^{ij}) - (\delta p_j)^- (x_n^{ij})=0 \label{onthecut} \ .
\end{equation}

On top of that, the quasi-momenta also present poles on the same points as the Lax connection. From the residue of the Lax connection we infer that
\begin{align}
	\frac{1}{s} \res_{x=s} \delta \hat{p}_1^A &= \frac{1}{s} \res_{x=s} \delta \hat{p}_1^S=-s\res_{x=1/s} \delta \check{p}_1^A=-s\res_{x=1/s} \delta \check{p}_1^S \ , \notag \\
	\frac{1}{s} \res_{x=s} \delta \hat{p}_2^A &= \frac{1}{s} \res_{x=s} \delta \hat{p}_2^S=-s\res_{x=1/s} \delta \check{p}_2^A=-s\res_{x=1/s} \delta \check{p}_2^S \ , \notag \\
	-s \res_{x=-1/s} \delta \hat{p}_1^A &= -s \res_{x=-1/s} \delta \hat{p}_1^S=\frac{1}{s}\res_{x=-s} \delta \check{p}_1^A=\frac{1}{s} \res_{x=-s} \delta \check{p}_1^S \ , \notag \\
	-s \res_{x=-1/s} \delta \hat{p}_2^A &= -s \res_{x=-1/s} \delta \hat{p}_2^S=\frac{1}{s}\res_{x=-s} \delta \check{p}_2^A=\frac{1}{s} \res_{x=-s} \delta \check{p}_2^S \label{correlatedresidues} \ .
\end{align}


These restrictions provides us with enough information to completely fix the corrections to the quasi-momenta. The details of the construction are lengthy, so we have relegated them to Appendix~\ref{Appendix}. We eventually obtain the following expression for the correction to the dispersion relation
\begin{align}
	w_0 \delta \Delta &=\sum_n{ \left[\left( n \kappa \hat{x}_n^{AA} -qn -w_0 \right) \hat{N}^{AA}_n  +\left( \frac{n \kappa}{\check{x}_n^{AA}}-qn \right) \check{N}^{AA}_n + \left( \frac{(n+2m)\kappa}{\check{x}_n^{SS}} -qn \right)\check{N}_n^{SS}  \right. } \notag \\
	&+\left( n \kappa\hat{x}_n^{SS} -qn +2 K(0) \right) \hat{N}^{SS}_n + \left(\frac{(n+m) \kappa}{\check{x}_n^F}-qn \right) (\check{N}_n^{AS}+\check{N}_n^{SA}) \notag \\
	&\left. +\left( n \kappa \hat {x}_n^F -qn -K(0) - w_0 \kappa\right) (\hat{N}_n^{AS}+\hat{N}_n^{SA})  \vphantom{\left( \frac{\kappa(n+2m)}{\check{x}_n^{SS}} -qn \right)\check{N}_n^{SS}} \right] \label{finitegapcorrection}
\end{align} 
where we still have to implement the expressions for each pole. We can check that in the $q\rightarrow 0$ limit we recover the undeformed $AdS_3\times S^3 \subset AdS_5\times S^5$ expressions \cite{BSZ}. Note that all $-2qn$ terms in the expressions from appendix~\ref{Appendix} have been replaced by $-qn$ terms to facilitate a later comparison with the frequencies obtained through the quadratic fluctuations. This substitution is legitimated by the level matching condition, expressed as
\begin{equation}
	\sum_n{n\sum_{\text{all exc.}}{N_n}}=0 \ .
\end{equation}
It is also convenient to define the frequencies $\Omega^i_n$ as
\begin{align}
	w_0 \delta \Delta &= \sum_n \left[ (\hat{\Omega}^{AA}_n -w_0 ) \hat{N}^{AA}_n + \check{\Omega}_n \check{N}^{AA}_n +\left( \hat{\Omega}_n^{SS} +2 K(0) \right) \hat{N}^{SS}_n +\check{\Omega}_n^{SS} \check{N}_n^{SS} \right. \notag \\
	&+\left. (\hat{\Omega}^F_n -K(0) -\kappa w_0) (\hat{N}_n^{AS}+\hat{N}_n^{SA}) + \check{\Omega}^F_n (\check{N}_n^{AS}+\check{N}_n^{SA}) \right] \label{definitionOmega} \ .
\end{align} 

The value of the poles is determined by equation (\ref{onthecut}), which can be solved as a series in ${\Upsilon^{-1}}$ for general values of $q$. Nevertheless, when taking the $q\rightarrow 1$ limit, those equations heavily simplify and we can find exact solutions after an appropriate regularization the poles with $\kappa$ factors. The solutions for general $q$ are collected in the appendix~\ref{generalq}. Here we just write down the solutions for $q=1$, which read
\begin{alignat}{2}
	\kappa \hat{x}^{AA}_{q\rightarrow 1} &=2\frac{n+\Upsilon}{n} \ , & \quad \quad \frac{\check{x}^{AA}_{q\rightarrow 1}}{\kappa} &=\frac{n}{2(n-\Upsilon)} \ , \notag \\
	\kappa \hat{x}^{SS}_{q\rightarrow 1} &=2\frac{n+\Upsilon}{n} \ , & \quad \quad \frac{\check{x}^{SS}_{q\rightarrow 1}}{\kappa} &=\frac{n+2m}{2(n+2m-\Upsilon)}  \ , \notag \\
	\kappa \hat{x}^{AS}_{q\rightarrow 1} &=2\frac{n+\Upsilon}{n} \ , & \quad \quad \frac{\check{x}^{AS}_{q\rightarrow 1}}{\kappa} &=\frac{n+m}{2(n+m-\Upsilon)} \ .
\end{alignat}
The $\kappa$ factors cancel after substituting into the one-loop correction (\ref{finitegapcorrection}), revealing that the limit is well behaved despite the apparent singularities that appear. Plugging back into the aforementioned equation and using the definitions (\ref{definitionOmega}) we get
\begin{alignat}{2}
	\hat{\Omega}^{AA}_{q\rightarrow 1} &=n+2\Upsilon \ , & \quad \quad \check{\Omega}^{AA}_{q\rightarrow 1} &=n-2\Upsilon \ , \notag \\
	\hat{\Omega}^{SS}_{q\rightarrow 1} &=n+2\Upsilon \ , & \quad \quad \check{\Omega}^{SS}_{q\rightarrow 1} &=n+4m-2\Upsilon \ , \notag \\
	\hat{\Omega}^{F}_{q\rightarrow 1} &=n+2\Upsilon \ , & \quad \quad \check{\Omega}^{F}_{q\rightarrow 1} &=n+2m-2\Upsilon \ . \label{algebraiccurvefrequencies}
\end{alignat}

We end this section comparing the expressions of the characteristic frequencies obtained through both methods and discuss their differences. Here we focus on the $\mathfrak{su}(2)$ sector in the pure NS-NS limit. The comparison for general values of $q$ is relegated to the appendix~\ref{generalq}, but the arguments presented in the discussion below are still valid.

When we collate equations (\ref{Fre}), (\ref{cuen}) and (\ref{cias}) with (\ref{algebraiccurvefrequencies}) we observe that they are equal up to some shifts. Such shifts fall into two categories, shifts of the mode number and shifts of the frequencies, and can be understood as a change of reference frame \cite{FinitegapGromov}. As our frequencies present the same shift structure and the shifts at $q=1$ cancel each other when summed, we confirm that both computations are in agreement. Therefore, we can extract the one-loop correction to the dispersion relation using the frequencies from either of the methods.

\section{Computation of the one-loop correction}
\label{comparison}

In this section we put together the characteristic frequencies from previous sections to compute the one-loop shift to the dispersion relation in the $\mathfrak{su}(2)$ sector. This correction is given by the sum of the fluctuation frequencies
\begin{equation}
	E_{\text{1-loop}} =E_0 +\delta E \ ,  \quad \quad \delta E=\frac{1}{2w_0} \sum_{n\in \mathbb{Z}}{( \omega_n^B - \omega_n^F)} \ ,
\end{equation}
where $\omega_n^B$ and $\omega_n^F$ are the bosonic and fermionic contributions respectively.

Firstly, we consider the pure NS-NS the limit in the $\mathfrak{su}(2)$ sector. Using the frequencies obtained from the quadratic fluctuations, these contributions are
\begin{align}
	\omega_n^B &= 2n+(n+w_0)+(n-w_0)+4n=8n \ , \notag \\
	\omega_n^F &= 2\left[ 2\left( n+ \frac{w_0}{2} \right) +2\left( n- \frac{w_0}{2} \right) \right]=8n \label{frequenciessummed} \ .
\end{align}
The net contribution from the frequencies thus vanishes, resulting in the vanishing of the one-loop correction in the pure NS-NS limit
\begin{equation}
\label{Vanishing}
	\delta E  \xrightarrow{q\rightarrow1}0 \ .
\end{equation}
This vanishing remains valid out of the $\mathfrak{su}(2)$ sector, i.e. if $m_1 \neq -m_2$.

Let us focus now on the case of general mixed flux. Here
\begin{align}
	\omega_n^{AdS} &= \sqrt{n^2 +2qn w_0 +w_0^2} +\sqrt{n^2 -2qn w_0 +w_0^2} \notag \\
	&=\sqrt{(n+qw_0)^2 +\kappa^2 w_0^2}+\sqrt{(n-qw_0)^2 +\kappa^2 w_0^2} \label{adssumed} \ ,\\
	\omega_n^{T} &=4n \label{tsumed} \ , \\
	\omega_n^{F} &=4n+ 4\sqrt{n^2 -q^2 w_0^2+\Upsilon^2} \label{fermionssumed} \ ,
\end{align}
whereas the $S^3$ contribution is more conveniently written after summing part of the series in a square root, by analogy with the results we have obtained from the algebraic curve,
\begin{align}
	\omega_n^S &=\left( \Upsilon +\sqrt{n^2 -2nq\Upsilon +\Upsilon^2} +\mathcal{O} \left( \Upsilon^{-2} \right) \right) + \left( -\Upsilon +\sqrt{n^2 +2nq\Upsilon +\Upsilon^2} +\mathcal{O} \left( \Upsilon^{-1} \right) \right) \notag \\
	&=\sqrt{(n+q\Upsilon)^2 +\kappa^2 \Upsilon^2}+\sqrt{(n-q\Upsilon)^2 +\kappa^2 \Upsilon^2} +\mathcal{O} \left( \Upsilon^{-1} \right)   \ . \label{ssumed}
\end{align}

In order to perform the infinite sum on the mode number we use the method presented in \cite{Nameki}, which consists in replacing such sum by an integral weighed with a cotangent function
\begin{equation}
\label{Sum}
2\pi i \sum_{n\in \mathbb{Z}} \omega_n= \oint_{\mathcal{C}} dz \, \pi \cot (\pi z) \omega_z \ ,
\end{equation}
where the contour $\mathcal{C}$ encircles the real axis.

The frequencies related to the $AdS$ space (\ref{adssumed}) and fermions (\ref{fermionssumed}) contain square roots with complex branch points. The same applies to the leading order of the frequencies related to $S^3$ (\ref{ssumed}). Choosing our branch cuts from each branch point to infinity allow us to deform the contour of the integral so it encircles them. As the branch points are of order $i\Upsilon$, we can consistently approximate the cotangent by one in the semiclassical limit, reducing the contour integral to the usual integral of a square root. In our case, (\ref{Sum}) involves integrals of the form
\begin{align}
\label{Integral}
	I (a,b)&=-\int_{a+ib}^{i\Lambda} dz \sqrt{(z-a)^2+b^2}=\int_{b-ia}^\Lambda{\sqrt{(z+ia)^2 -b^2}} \notag \\
	&= \frac{1}{4} \left[ -2\Lambda^2 -4ia\Lambda-2a^2 +b^2  (1+\log 4) -2b^2 (\log b-\log \Lambda ) \right] +\mathcal{O} ( \Lambda^{-1} )\ ,
\end{align}
where $\Lambda$ is a sharp cut-off which regularizes (\ref{Integral}). The sum of these contributions gives
\begin{align}
	&\sum_{n\in \mathbb{Z}} \left( \omega_n^S +\omega_n^{AdS} +\omega_n^{T} - \omega_n^{F} \right) = \left[ I(q \Upsilon , \kappa \Upsilon) +I(-q \Upsilon , \kappa \Upsilon)+ I (q w_0, \kappa w_0) +I (-q w_0, \kappa w_0) \vphantom{\left( 0,\sqrt{\Upsilon^2 -q^2 w_0^2} \right))} \right. \notag \\
	&\left.-4 I \left( 0,\sqrt{\Upsilon^2 -q^2 w_0^2} \right) \right] \label{schafer} \ .
\end{align}
A direct inspection shows that the quadratic and linear contributions of the regulator cancel but the logarithmic contribution does not. This fact reflects that higher orders in the expansion in the mode number of the $S^3$ frequencies contribute to the cancellation. In order to check this statement we should expand the characteristic frequencies of the $S^3$ modes in the regime of large mode number $n$
\begin{align}
	\omega^{S}_{1,n} &=n+(1-q) w_0+\frac{\kappa^2 [\Upsilon^2-m^2 (1+q^2)]}{2n} +\frac{\kappa^2 [m^4 +q (w_0 -m^2)^2]}{2 w_0 n^2} +\mathcal{O} (n^{-3}) \ , \notag \\
	\omega^{S}_{2,n} &=n-(1-q) w_0+\frac{\kappa^2 [\Upsilon^2-m^2 (1+q^2)]}{2n} -\frac{\kappa^2 [m^4 +q (w_0 -m^2)^2]}{2 w_0 n^2} +\mathcal{O} (n^{-3}) \ , \label{largeN}
\end{align}
where we have written down just the solutions with $+n$ as leading contribution. Since matching with the frequencies expanded in $\Upsilon$ is not direct, the labelling here is arbitrary. Taking into account the relation
\begin{equation}
	2 \kappa^2 [\Upsilon^2-m^2 (1+q^2)] = 4 (\Upsilon^2 -q^2 w_0^2)-2\kappa^2 w_0^2 \ ,
\end{equation}
we can check that after replacing the first two contributions in (\ref{schafer}) by the sum over $n$ of the frequencies (\ref{largeN}), the logarithmic divergence now cancels.  This proves that the one-loop correction to the dispersion relation is finite for all values of the mixing parameter.

\section{A comment about non-rigid strings}
\label{non-rigid}

In this section we provide a plausibility argument for the vanishing of the one-loop correction for the non-rigid case in the pure NS-NS limit.

First we must summarize some results about the non-rigid spinning string. The existence of an Uhlenbeck constant of motion allows us to reduce the equations of motion (\ref{r1prime}), (\ref{r2prime}) and (\ref{alphaprime}) into a first-order differential equation. This equation is solved in terms of Jacobi elliptic functions, giving
\be
r_1^2(\sigma) = c_1+c_2 \, \hbox{sn}^2
\big( c_3 \sigma , \nu \big) \ ,
\label{r1elliptic}
\ee
where $c_i$ and $\nu$ are constants that depend on $\omega_i$, $v_i$ and $q$ whose explicit expressions can be found in \cite{HN2}. For our purposes it is enough to know that the elliptic parameter $\nu$ vanishes when $q\rightarrow 1$. Another important feature of the solution (\ref{r1elliptic}) is that its functional form is the same as the one for non-rigid spinning strings in $AdS_5 \times S^5$ \cite{NR}.

The one-loop corrections for spinning folded and pulsating strings in $AdS_5 \times S^5$ were computed in \cite{dunne,Forini}, where it was shown that the Euler-Lagrange equations of all the fluctuations can be rewritten as the eigenvalue problem of a single-gap Lamé operators
\begin{equation}
	[ \partial_x^2 + 2 \bar{\nu}^2 \text{sn}^2 (x|\bar{\nu}) +\Omega^2 ] f_\Lambda (x)= \Lambda f_\Lambda (x) \ .
\end{equation}
Here $\bar{\nu}$ is related to the elliptic modulus of the Jacobi function involved on the classical solution, $x$ is a linear function of $\sigma$ for spinning strings and $\Omega^2$ is a linear function of the square of the characteristic frequency. The specific relations depend on which of the fluctuations are we considering, but in all the cases $\bar{\nu}$ vanishes if the elliptic modulus of the classical solution vanishes.

Since $\nu$ vanishes in the limit $q=1$, all the Lamé equations reduce to a wave equation in such limit as long as the functional form of this eigenvalue problem remains unaltered when we include the NS-NS flux. Accordingly, the characteristic frequencies should become a linear function of the mode number $n$ and the one-loop correction would vanish in the non-rigid case.

\section{Conclusions}
\label{conclusions}

In this article we have derived the one-loop correction to the dispersion relation of rigid closed spinning strings on the $\mathbb{R}\times S^3$ subsector of a type IIB string theory on  $AdS_3 \times S^3 \times T^4$ background supported by a mixture of R-R and NS-NS fluxes. This correction has been obtained through the computation of the characteristic frequencies via both background field expansion and the finite-gap equations without massless excitations. We have proved that the correction remains finite for all values of the parameter that controls the mixing of the two fluxes. In addition, all the frequencies can be computed analytically in the pure NS-NS limit, where they become linear and the one-loop correction vanishes. We have also argued that this vanishing probably extends to non-rigid strings.

It is important to remark that our finite-gap computation does not take into account the massless fields because the background field expansion showed that their net contribution vanishes. Including the massless contributions into the finite-gap equations would need a further extension of the procedure followed here. It would be desirable to generalize the method presented in \cite{masslessfinitegap} to deal with these massless excitations in the case of mixed flux.

A natural generalization of our analysis would be the precise computation of the one-loop correction for non-rigid strings. Even though the vanishing of this correction on the $q\rightarrow 1$ limit seems plausible, an explicit check is needed. In principle, the procedure used in \cite{dunne,Forini} for $AdS_5 \times S^5$ string theory could be generalized to this end.

Semiclassical giant magnon solutions support this statement, as they can be obtained as a particular regime of general spinning strings. These solutions were studied in \cite{B.Hoare,C.Ahn}, where it was shown that they display a linear dispersion relation. \footnote{Note that in \cite{C.Ahn} a more general magnonic dispersion relation has been derived. Nonetheless, semiclassical solutions can be typically mapped via AdS/CFT duality in the $J_{i}\rightarrow \infty$ limit, according to which the dispersion relation therein indeed becomes linear.} Furthermore, in \cite{B.Hoare} it is argued that such dispersion relation holds at each perturbative order for giant magnons understood as magnonic bound states, which in particular implies the vanishing of its one-loop correction up to corrections in the coupling constant $h$. The construction of the S-matrix for elementary magnonic excitations starting from symmetry considerations strongly supports this fact \cite{completeworldsheet}. Equation (\ref{Vanishing}) and the discussion of section 6 suggests that the vanishing of the one-loop correction in the pure NS-NS limit is not a characteristic of rigid spinning string solutions but it might be a feature of general spinning strings. Proving this statement would shed light on the role of spinning semiclassical solutions in the mixed flux scenario, and its $q\rightarrow 1$ limit.

Besides, it would be desirable to compare our results with the prediction from the string Bethe ansatz for the dressing phase. The comparison is based on the realization of rigid spinning string solutions in the $\mathfrak{su}(2)$ sector from the point of view of the Bethe equations, along the lines of Hernández-López construction \cite{HernandezLopez}. In particular, such realization involves a large quantity of $\mathfrak{su}(2)$ Bethe roots lying in a single connected curve of the complex plane. The construction shows that there are two possible sources of corrections: quantum corrections to the classical integrable structure and wrapping corrections coming from finite size effects. In the $AdS_{5}\times S^5$ background, the comparison allowed to extract strong-coupling corrections to the dressing phase that later were proven to be in agreement with the predictions derived from crossing relations. In contrast to the $AdS_{5}\times S^5$ comparison, in $AdS_3  \times S^3 \times T^4$ with pure R-R flux both computations of the dressing phase did not match. The presence of massless modes, with no analogy in $AdS_5 \times S^5$, is believed to underlie this disagreement as wrapping contributions are exponentially supressed by the mass of the excitations involved.~\footnote{Perturbative analyses around the BMN vacuum of the dressing phase in the R-R \cite{Aniceto} and mixed flux \cite{SW2} regimes has also led to discrepancies with the all-loop S-matrix predictions and the dispersion relation, obtained both from the underlying symmetries of the system, when dealing with massless excitations.} On the other hand, it has been recently shown \cite{M.Baggio,A.Dei} that in the pure NS-NS limit wrapping corrections are absent and both the S-matrix and the string Bethe equations present a remarkably simple structure \cite{M.Baggio,A.Dei}. \footnote{It is not obvious how to apply the pure NS-NS limit to the flux-deformed Bethe equations proposed in \cite{Ads3S3T4A}, based on the centrally extended light-cone symmetry algebra. This is so because such algebra cannot be centrally extended when $q=1$.} Taking all of this into account, we expect that some control over massless wrapping corrections in the Hernández-López construction can be gained by means of the parameter $q$, helping to elucidate the disagreement.

Finally, we want to point out that the spectrum of closed strings with zero winding and zero momentum on the torus on a F1/NS5-brane supported with R-R moduli has been studied in \cite{Modulispace}. It is also shown there that the mixed flux background studied here can be retrieved from the near-horizon geometry of such scenario. Thus, our computation may also be relevant to the pure NS-NS theory at a generic point in its moduli space.

\section*{Acknowledgments}

The authors want to thank Rafael Hernández, Alessandro Torrielli and Santiago Varona for valuable comments on the manuscript. We also want to thank Ben Hoare for feedback on this work and for pointing us the articles \cite{dunne}. J. M. N. also want to thanks the Trinity College Dublin for its hospitality during the ``Workshop on higher-point correlation functions and integrable AdS/CFT''.

\appendix

\section{Conventions}
\label{conventions}

In this appendix we collect the conventions we used during the article, in particular those concerning the fermionic Lagrangian (\ref{Fermionic Lagrangian density}).

Firstly, we fix our index notation. We use greek indices for the worldsheet coordinates, lower-case latin indices for the ten-dimensional Minkowski flat spacetime, upper-case undotted indices for target spacetime and upper-case dotted indices separate the $32$ components Majorana-Weyl spinors in 10 dimensions into two $16$ components spinors.

The worldsheet coordinates, denoted as $\tau$ and $\sigma$, are raised and lowered with the flat metric $\eta_{\alpha \beta}=$diag$(-1,1)$ and its associated Levi-Civita symbol is defined so $\epsilon^{\tau \sigma}=1$. Flat Minkowskian coordinates takes values in $\{0,\dots ,9\}$, being raised and lowered with the flat metric $\eta_{ab}=$diag$(-1,1,\dots ,1)$.

In order to relate target space coordinates and flat Minkowskian coordinates we construct the vielbeins $E^{a}=E_{A}^{a} dX^A$ and the spin connection differential 1-forms $\Omega_{ab}=\Omega_{cab} E^{c}$, obtained from the former using
\begin{equation}
\d E^{a}+{\Omega^{a}}_{b}\wedge E^{b}=0 \ .
\end{equation}

For the $AdS_3 \times S^3$ space the vielbeins are given by
\begin{align}
	E^{0}&=z_0 \text{ d}t \ , & E^{1}&=-z_0\text{ d}z_1 +z_1\text{ d}z_0 \ , & E^{2} &=z_1 \text{ d}\phi \ , \nonumber \\
	E^{3}&=-r_1 \text{ d}r_2 +r_2 \text{ d}r_1 \ , & E^{4}&=r_1 \text{ d}\varphi_{1} \ , & E^{5}&=r_2 \text{ d}\varphi_{2} \ ,
\end{align}
The spin connection differential one-forms are written down as
\begin{align}
	\Omega_{01} &=-z_1 \text{ d} t \ , & \Omega_{21} &=z_0 \text{ d} \phi \ , &\Omega_{43} &=r_2 \text{ d} \varphi_1 \ , & \Omega_{53} &=-r_1 \text{ d} \varphi_{2} \ ;   
\end{align}
the remaining components either could be obtained through ${\Omega^{a}}_{b}=-{\Omega^{b}}_{a}$, or are zero. Both vielbeins and spin connection can be pulled back to the worldsheet using a solution of the equations of motion as
\begin{align*}
	e^{a}_{\alpha} &=E^{a}_{A}\partial_{\alpha}X^{A} \ , &  \omega_{\alpha ab} &= e^{c}_{\alpha}\Omega_{cab} \ .
\end{align*}
When the constant radii classical solution presented in section~\ref{classicalsetting} is plugged, the non-trivial pulled back vielbiens are 
\begin{align}
	e^{0} &=\kappa\textnormal~{d}\tau \ , & e^{4}&=a_{1}(\omega_{1}\textnormal{d}\tau+\alpha_{1}'\textnormal{d}\sigma) \ , & e^{5}&=a_{2}(\omega_{2}\textnormal{d}\tau+\alpha_{2}'\textnormal{d}\sigma) \ ,
\end{align}
and the non-trivial pulled-back spin connection differential one-forms are
\begin{align}
	\omega_{43} &=a_{2}(\omega_{1}\d\tau+\alpha'_{1}\d\sigma) \ , & \omega_{53}&=-a_{1}(\omega_{2}\d\tau+\alpha_{2}'\d\sigma) \ .    
\end{align}

$H_{abc}$ and $F_{abc}$ refer the Minkowskian components of the Neveu-Schwarz-Neveu-Schwarz and Ramond-Ramond three-form fluxes respectively, given by \cite{completeworldsheet}
\begin{align*}
	\slashed{H}_{a} &=2q \left[ \slashed{E}_a (\Gamma^{012} +\Gamma^{345})+ (\Gamma^{012} +\Gamma^{345}) \slashed{E}_a \right] \ , \\
	F_{abc} &= 12\kappa (\Gamma^{012} +\Gamma^{345}) \ ,
\end{align*}
where the slash denotes contraction with the gamma matrices. Integrability and conformal symmetry fix $q^2+\kappa^2=1$ \cite{CagnazzoZarembo}.

\section{Computation of the corrections to the quasi-momenta of the one-cut solution}
\label{Appendix}

We treat the $AdS_3$, the $S^3$ and the fermionic contributions separately to simplify the computation. This separation allow us to alleviate notation by dropping the sheet labels both on the poles $x$ and on the number of excitations/cuts $N_n$.

\subsection{Contribution from $AdS_3$ excitations}

The classical algebraic curve presents a cut only on the sheets related to the sphere, making the process of computing the $AdS_3$ modes equivalent to the computation of fluctuations around the BMN string solution. This computation has been already performed in \cite{FinitegapBabichenko}, so we just quote their result here (without imposing the level matching condition)
\begin{align}
	\delta \Delta &= 2\sum_n{ \left[ \hat{N}_n \left( h \kappa \hat{\alpha} (\hat{x}_n)-1-\frac{2qn}{w_0} \right) + \check{N}_n \left( \frac{h\kappa \check{\alpha} (\check{x}_n)}{\check{x}_n^2} -\frac{2qn}{w_0} \right)\right]} \ .
\end{align}
Substituting the quasimomenta (\ref{adsquasimomenta}) on the cut condition (\ref{onthecut}) relates the functions $\hat{\alpha}$ and $\check{\alpha}$ with the mode number and the position of the poles
\begin{align}
	\hat{p}_1^A (\hat{x}_n)-\hat{p}_2^A (\hat{x}_n) &=2 \frac{2\pi \hat{x}_n w_0}{\kappa (\hat{x}_n-s) \left( \hat{x}_n+\frac{1}{s} \right)}=\frac{4\pi h \hat{\alpha} (\hat{x}_n) w_0}{\hat{x}_n}=2\pi n \ , \\
	\check{p}_2^A (\check{x}_n)-\check{p}_1^A (\check{x}_n) &=2\frac{2\pi \check{x}_n w_0}{\kappa (\check{x}_n+s) \left( \check{x}_n-\frac{1}{s} \right)}=\frac{4\pi h \check{\alpha} (\check{x}_n) w_0}{\check{x}_n}=2\pi n \ .
\end{align}
Plugging them into the previous expression we obtain
\begin{align}
	\delta \Delta &= \frac{1}{w_0} \sum_n{ \left[ \hat{N}_n \left( n \kappa \hat{x}_n -2nq-w_0 \right) +\check{N}_n \left( \frac{n \kappa}{\check{x}_n}-2nq \right) \right] } \ . 
\end{align}

\subsection{Contribution from $S^3$ excitations}

Adding infinitesimal cuts in the sheets related to $S^3$ has to take into account the existence of a branch cut on the classical algebraic curve. The presence of both cuts generates two kinds of corrections to the quasi-momenta, one coming from the poles associated to the infinitesimal cuts and another coming from shifts of the branch points of the cut due to the addition of these poles. Thus we have to split the ansatz for the corrections to the quasi-momenta in two contributions
\begin{equation}
	\delta \check{p}^S_2=f(x)+\frac{g(x)}{K(x)} \ . \label{S3ansatz}
\end{equation}
where the factor $K(x)$ dividing the second term comes from the fact that $\partial_{x_0} \sqrt{x-x_0}\propto 1/\sqrt{x-x_0}$. Using both the analytic properties of the corrections to the quasi-momenta on the cut (\ref{onthecut}) and the inversion symmetry of the algebraic curve 
\be
\delta \hat{p}_i^S (x)=\delta \check{p}_i^S \left( \frac{1}{x} \right) \ ,
\ee
we can fix the rest of the quasi-momenta related to the sphere
\begin{align}
	\delta \check{p}^S_1 (x) &=f(x)-\frac{g(x)}{K(x)} \ , &	\delta \check{p}^S_2 (x) &=f(x)+\frac{g(x)}{K(x)} \ , \notag \\
	\delta \hat{p}^S_1 (x) &=f\left( \frac{1}{x} \right)-\frac{g\left( \frac{1}{x} \right)}{K\left( \frac{1}{x} \right)} \ , &	\delta \hat{p}^S_2 (x) &=f\left( \frac{1}{x} \right)+\frac{g\left( \frac{1}{x} \right)}{K\left( \frac{1}{x} \right)} \ .
\end{align}

The explicit form of these functions is obtained from the known pole structure of the corrections and their asymptotic properties. Equation~(\ref{checkpoles}) entails that the combinations $\delta \check{p}^S_1+\delta \check{p}^S_2$ and $\delta \hat{p}^S_1+\delta \hat{p}^S_2$ have no poles, hence we are free to choose $f(x)=0$. On the other hand, $\delta \check{p}^S_2-\delta \check{p}^S_1$ has a simple pole with residue $2\check{\alpha} (\check{x}) \check{N}_n$ at $\check{x}$ and $\delta \hat{p}^S_2-\delta \hat{p}^S_1$ has a simple pole with residue $2\hat{\alpha} (\hat{x}) \hat{N}_n$ at $\hat{x}$. Furthermore, $\delta \check{p}^S_i$ (respectively $\delta \hat{p}^S_i$) have poles at $-s$ and $\frac{1}{s}$ (respectively $s$ and $-\frac{1}{s}$) whose residues are correlated with those of the $\delta \check{p}^A_{i}$  (respectively $\delta \hat{p}^A_{i}$) quasi-momenta at the same points, see (\ref{correlatedresidues}). As a consequence, we can write down the following ansatz for the function $g(x)$
\begin{align}
	g(x) &=a+\frac{s \delta a_1}{x+s}+\frac{\delta a_2}{sx-1}+ \sum_n \left( \check{N}_n \frac{\check{\alpha} (\check{x}_n) K(\check{x}_n)}{x-\check{x}_n}-\hat{N}_n\frac{\hat{\alpha} (\hat{x}_n) K(1/\hat{x}_n)}{\hat{x}_n (1-x\hat{x}_n)} \right) \\
	&=a+\frac{-2 \kappa \delta a_- +2x(\delta a_+ +q \delta a_-)}{\kappa (x+s)\left( x-\frac{1}{s} \right)}+ \sum_n \left( \check{N}_n \frac{\check{\alpha} (\check{x}_n) K(\check{x}_n)}{x-\check{x}_n}-\hat{N}_n\frac{\hat{\alpha} (\hat{x}_n) K(1/\hat{x}_n)}{\hat{x}_n (1-x\hat{x}_n)} \right) \ , \notag
\end{align}
where $a$ and $\delta a_i$ are unknown constants.

As we mentioned in section~\ref{algebraiccurve}, our conventions for the definition of the Lax connection allow us to relate it with the Noether currents of the system for large values of the spectral parameter in a simple way. Therefore, the asymptotic behaviour of the quasi-momenta can be related with conserved global charges. In particular, for the excitations we are interested in we have
\begin{align}
	\lim_{x\rightarrow \infty} \kappa h x \, \delta \check{p}^S_1 (x) & =-\check{N} \ , & \lim_{x\rightarrow \infty} \kappa h x \, \delta \check{p}^S_2 (x) & =\check{N} \ , \\
	\lim_{x\rightarrow \infty} \kappa h x \, \delta \hat{p}^S_1 (x) & =\hat{N} \ , & \lim_{x\rightarrow \infty} \kappa h x \, \delta \hat{p}^S_2 (x) & =-\hat{N} \ ,
\end{align}
with $N=\sum_n N_n$. These asymptotic properties are mapped to asymptotic properties of $g(x)$ at $x\rightarrow\infty$ and $x\rightarrow 0$. The behaviour at infinity fixes the constant $a$ to
\begin{equation}
	\lim_{x\rightarrow \infty} g(x)=a=\lim_{x\rightarrow \infty} \frac{K (x)}{2} (\delta \check{p}_2 (x) - \delta \check{p}_1 (x))=\frac{m \check{N}}{h} \ .
\end{equation}
Besides, the behaviour around zero provides us two conditions that fix the remaining two unknown constants
\begin{gather}
	\frac{g(x)}{K(x)} = \delta \hat{p}_2 (1/x) - \delta \hat{p}_1 (1/x) \ , \notag \\
	 \frac{g(0)}{K(0)} + \left( \frac{g'(0)}{K(0)} - \frac{g(0)}{K(0)} \, \frac{K'(0)}{K(0)} \right) x +\mathcal{O} (x^2) = -\frac{\hat{N}}{\kappa h} x +\mathcal{O} (x^2) \label{asymptoticzero}\ .
\end{gather}

To simplify our results we write the residues $\hat{\alpha}$ and $\check{\alpha}$ in terms of the poles using  the equation~(\ref{onthecut}) and the classical values of the quasi-momenta (\ref{shatquasimomenta}) and (\ref{scheckquasimomenta})
\begin{align}
	\hat{p}^S_1 (\hat{x}_n)-\hat{p}^S_2 (\hat{x}_n) &=2\pi n \Rightarrow 2\frac{ \hat{\alpha} (\hat{x}_n) h K(1/\hat{x}_n)}{\hat{x}_n}=n \ , \label{poleeq1} \\
	\check{p}^S_2 (\check{x}_n) -\check{p}^S_1 (\check{x}_n) &=2\pi n \Rightarrow 2\frac{ \check{\alpha} (\check{x}_n) h K(\check{x}_n)}{\check{x}_n}=n+2m \label{poleeq2} \ .
\end{align}
The $\mathcal{O}(1)$ of (\ref{asymptoticzero}) reads
\begin{equation}
	g(0)=a+2\delta a_- -\sum_n \left( \frac{\hat{N}_n n}{2h} + \frac{\check{N}_n (n+2m)}{2h} \right)=0 \Rightarrow \delta a_-=\sum_n  \frac{(\hat{N}_n+\check{N}_n) n}{4h} \ ,
\end{equation}
while the $\mathcal{O} (x)$ gives us
\begin{equation}
	\delta a_+ -q \delta a_- =-\sum_n \frac{\kappa}{4h} \left[ \frac{n+2m}{\check{x}_n} \check{N}_n + \left( n \hat{x}_n -\frac{2 K(0)}{\kappa} \right) \hat{N}_n \right] \ .
\end{equation}

Now that we know the residues at $-s$ and $1/s$ of the quasimomenta associated to the sphere, we can compute the correction to, for example, the $\check{p}_2^A$ quasi-momenta, as $\delta\Delta$ is computed from its asymptotic behaviour. Using that $K(-s)=K(1/s)=w_0$ we can write
\begin{equation}
	\delta \check{p}_2^A (x)=\frac{-2 \kappa \delta a_- +2x(\delta a_+ +q \delta a_-)}{w_0 \kappa (x+s)\left( x-\frac{1}{s} \right)} \ ,
\end{equation}
and thus
\begin{align}
	-\frac{\delta \Delta}{2\kappa h} &=\lim_{x\rightarrow \infty} x \, \delta \check{p}_2^A (x)=\frac{2(\delta a_+ +q \delta a_-)}{w_0 \kappa} \notag \ , \\
	\delta \Delta &=\sum_n\frac{\kappa}{w_0} \left[ \left( \frac{n+2m}{\check{x}_n}  -2qn \right) \check{N}_n + \left( n \hat{x}_n -2qn-\frac{2 K(0)}{\kappa} \right) \hat{N}_n \right] \ .
\end{align}

\subsection{Contribution from fermionic excitations}

As opposed to the previous cases, we have to distinguish between the fermionic contributions $N^{AS}$ and $N^{SA}$, both hatted and checked. Nevertheless, equations (\ref{adsquasimomenta}), (\ref{shatquasimomenta}) and (\ref{scheckquasimomenta}) imply that the differences of classical quasi-momenta are equal two by two
\begin{align}
	\hat{p}_1^A (x)-\hat{p}_2^S (x) &=-\hat{p}_2^A (x) +\hat{p}_1^S (x) \ , & \check{p}_2^A (x) -\check{p}_1^S (x) &=-\check{p}_1^A (x) +\check{p}_2^S (x) \ ,
\end{align}
hence both kind of poles are equal in pairs $\hat{x}^{AS}=\hat{x}^{SA}$ and $\check{x}^{AS}=\check{x}^{SA}$. 

As we have to deal with the cut of the classical solution too, we can use an analogue ansatz to the one we used for the $S^3$ modes (\ref{S3ansatz}), but with different functions $f(x)$ and $g(x)$ that reflect the pole structure of fermionic fluctuations (\ref{checkpoles}). The ansätze for such functions are
\begin{align}
	f(x) &=a+\frac{2(\delta a_+ +q\delta a_-) x -2\kappa \delta a_-}{\kappa (x+s) \left( x-\frac{1}{s} \right)} +\sum_n \left( \frac{\check{\alpha} (\check{x}_n)}{x-\check{x}_n} \check{n}_n -\frac{\hat{\alpha} (\hat{x}_n)}{\hat{x}_n(1-x \hat{x}_n)} \hat{n}_n \right) \ , \\
	g(x) &=b+\frac{2(\delta b_+ +q\delta b_-) x -2\kappa \delta b_-}{\kappa (x+s) \left( x-\frac{1}{s} \right)} +\sum_n \left( \frac{K (\check{x}_n) \check{\alpha} (\check{x}_n)}{x-\check{x}_n} \check{N}_n -\frac{K(1/\hat{x}_n) \hat{\alpha} (\hat{x}_n)}{\hat{x}_n (1-x \hat{x}_n)} \hat{N}_n \right)
\end{align}
where we have defined $2\check{n}\equiv \check{N}^{AS}-\check{N}^{SA}$ and $2\check{N}\equiv \check{N}^{AS}+\check{N}^{SA}$, with similar expression for the hatted ones. Again, the known asymptotic behaviour of the quasi-momenta fixes the unknown constants. 

Let us start with the function $f(x)$
\begingroup
\allowdisplaybreaks
\begin{align}
	f(x)&= a +\left( \frac{2(\delta a_+ + q\delta a_-)}{\kappa} + \sum_n \check{\alpha} (\check{x}_n) \check{n}_n + \sum_n \frac{\hat{\alpha} (\hat{x}_n) \hat{n}_n}{\hat{x}_n^2} \right) \frac{1}{x}+\mathcal{O} \left( \frac{1}{x^2} \right) \notag \\
	&=\frac{\check{n}}{\kappa h x} +\mathcal{O} \left( \frac{1}{x^2} \right) \ , \\
	f(x)&= a +2\delta a_- -\sum_n \frac{\check{\alpha} (\check{x}_n) \check{n}_n}{\check{x}}-\sum_n \frac{\hat{\alpha} (\hat{x}_n) \hat{n}_n}{\hat{x}_n} +\left[ -\frac{2(\delta a_+ + q\delta a_-)}{\kappa}+2\delta a_- \left( s-\frac{1}{s} \right) \right. \notag \\
	&\left. - \sum_n\frac{\check{\alpha} (\check{x}_n) \check{n}_n}{\check{x}_n^2} -\sum_n \hat{\alpha} (\hat{x}_n) \hat{n}_n \right] x +\mathcal{O} (x^2) =-\frac{\check{n} x}{\kappa h} +\mathcal{O} (x^2) \ .
\end{align}
\endgroup
These are four equations for three unknowns. We can check that the equations are compatible and, in fact, solved by
\begin{align*}
	 a&=0 \ , \\
	2\delta a_- &=\sum_n \left( \frac{\check{\alpha} (\check{x}_n) \check{n}_n}{\check{x}_n} +\frac{\hat{\alpha} (\hat{x}_n) \hat{n}_n}{\hat{x}_n} \right) \ , \\
	-2\delta a_+ &=\sum_n\frac{\check{\alpha} (\check{x}_n) \check{n}_n}{\check{x}_n} \left( \frac{\kappa}{\check{x}_n} -q \right) +\sum_n \frac{\hat{\alpha} (\hat{x}_n) \hat{n}_n}{\hat{x}_n} \left( \frac{\kappa}{\hat{x}_n} -q \right) \ .
\end{align*}
Substituting back these expressions, $f(x)$ can be rewritten in a more compact form
\begin{equation}
	f(x)= \frac{x}{h \kappa (x+s) \left( x-\frac{1}{s} \right)} \sum_n \left( \frac{x \check{n}_n}{x-\check{x}_n}+\frac{\hat{n}_n}{1-x \hat{x}_n} \right) \ .
\end{equation}

In addition, the function $g(x)$ has the expansions
\begin{align}
	g(x)&= b +\mathcal{O} \left( \frac{1}{x} \right) =K(\infty) \frac{\check{N}}{\kappa h x} +\mathcal{O} \left( \frac{1}{x} \right)=\frac{m \check{N}}{h}+\mathcal{O} \left( \frac{1}{x} \right) \ , \\
	g(x)&= b +2\delta b_- -\sum_n \frac{\check{\alpha} (\check{x}_n) K(\check{x}_n) \check{N}_n}{\check{x}_n} -\sum_n \frac{\hat{\alpha} (\hat{x}_n) K(1/\hat{x}_n) \hat{N}_n}{\hat{x}_n}+\left[ -\frac{2(\delta b_+ + q\delta b_-)}{\kappa} \right. \notag \\
	&\left. +2\delta b_- \left( s-\frac{1}{s} \right)- \sum_n \frac{\check{\alpha} (\check{x}_n) K(\check{x}_n) \check{N}_n}{\check{x}_n^2} -\sum_n \hat{\alpha} (\hat{x}_n) K(1/\hat{x}_n) \hat{N}_n \right] x +\mathcal{O} (x^2) \notag \\
	&=-\frac{K(0) \check{N} x}{\kappa h} +\mathcal{O} (x^2) \ .
\end{align}
The computation of the explicit value of the unknown constants is simplified when the residues are rewritten in terms of the poles via eq.~(\ref{onthecut})
\begin{align}
	\check{p}_2^S (\check{x}_n)-\check{p}_1^A (\check{x}_n)=\frac{2 \pi h \check{\alpha}(\check{x}_n) K(\check{x}_n)}{\check{x}_n}+\frac{2 \pi \check{w}_n}{h}-m=2\pi n &\Rightarrow \frac{\check{\alpha}(\check{x}_n) K(\check{x}_n)}{\check{x}_n}=\frac{n+m-\check{w}_n}{h} \notag \\
	\hat{p}_1^A (\hat{x}_n) -\hat{p}_2^S (\hat{x}_n)=2\pi n &\Rightarrow \frac{\hat{\alpha}(\hat{x}_n) K(1/\hat{x}_n)}{\hat{x}}=\frac{n-\hat{w}_n}{h}
\end{align}
where we have defined $\check{w}_n\equiv \frac{\check{x}_n w_0}{\kappa (\check{x}_n+s) \left( \check{x}_n-\frac{1}{s} \right)}$ and $\hat{w}_n\equiv \frac{\hat{x}_n w_0}{\kappa (\hat{x}_n-s) \left( \hat{x}_n+\frac{1}{s} \right)}$. Therefore we can write the unknown coefficients as
\begin{align*}
	b&=\frac{m \check{N}}{h} \ , \\
	-2 &\delta b_- =\sum_n \left( \frac{\check{w}_n-n}{h}\check{N}_n+\frac{\hat{w}_n-n}{h}\hat{N}_n  \right)\ , \\
	&\delta b_+ =\frac{\kappa}{2h} \left[ \frac{2qh\delta b_-}{\kappa} +\sum_n \frac{\check{w}_n-n-m}{\check{x}_n} \check{N}_n +\sum_n \left( \frac{K(0)}{\kappa} +(\hat{w}_n-n)\hat{x}_n \right) \hat{N}_n \right] \ ,
\end{align*}

Now that we have reconstructed the $\delta \check{p}_i^S$ quasi-momenta, we can focus on the anti-de Sitter sheets. For example, we can construct $\delta \check{p}_2^A$ and extract $\delta \Delta$ from the comparison of the residues of the former at $-s$ and $1/s$ with the residues of $\delta \check{p}_2^A$. As the steps involved in the construction of $\delta \check{p}_2^A$ are similar to those in fixing $f(x)$, we can write down directly
\begin{equation}
	\delta \check{p}_2^A=\frac{x}{h \kappa (x+s) (x-\frac{1}{s})} \left( \frac{-\delta \Delta}{2} +\sum_n \frac{\hat{N}_n-\hat{n}_n}{1-x \hat{x}_n} +\sum_n \frac{x (\check{n}_n-\check{N}_n)}{x-\check{x}_n} \right) \ .
\end{equation}
Imposing
\begin{equation}
	\res_{x=-s} \delta \check{p}_2^A +\res_{x=1/s} \delta \check{p}_2^A =\res_{x=-s} \delta \check{p}_2^S +\res_{x=1/s} \delta \check{p}_2^S \ ,
\end{equation}
and using that $K(-s)=K(1/s)=w_0$, we get
\begin{multline*}
	-\frac{\delta \Delta}{h \kappa} +\sum_n \left[ \left( \frac{s}{1+s \hat{x}_n} +\frac{1}{s-\hat{x}_n} \right) \frac{\hat{n}_n-\hat{N}_n}{h \kappa \left( s+\frac{1}{s} \right)} +\left( \frac{s^2}{s+\check{x}_n} +\frac{1}{s(1+s\check{x}_n)} \right) \frac{\check{n}_n -\check{N}_n}{h \kappa \left( s+\frac{1}{s} \right)} \right] \\
	=\frac{1}{h \kappa \left( s+\frac{1}{s} \right)} \sum_n \left[ \left( \frac{s}{1+s \hat{x}_n} +\frac{1}{s-\hat{x}_n} \right) \hat{n}_n +\left( \frac{s^2}{s+\check{x}_n} +\frac{1}{s(1+s\check{x}_n)} \right) \check{n}_n \right] +\frac{2 (\delta b_+ +q \delta b_-)}{\kappa w_0} \ ,
\end{multline*}
which reduces to
\begin{equation}
	\delta \Delta = \frac{1}{w_0} \sum_n \left[ \left( \frac{n+m}{\check{x}_n/\kappa}-2qn \right) \check{N}_n +\left( n \kappa \hat {x}_n -K(0) -2qn-\kappa w_0 \right) \hat{N}_n \right] \ .
\end{equation}

\section{Finite gap frequencies for general values of $q$}
\label{generalq}

In this appendix we write down the frequencies computed using the finite-gap equations for general values of the mixing parameter $q$. The equations~(\ref{onthecut}) for the placement of the poles can be solved exactly for general $q$ for the $AdS_3$ and the fermionic fluctuations, but not for the $S^ 3$ fluctuations, which have to be expressed as a series on ${\Upsilon}^{-1}$.

The values for the poles involving first two kinds of excitations are
\begin{align}
	\hat{x}_n^{AA} &=\frac{nq+w_0 \pm \sqrt{n^2 +2n q w_0 +w_0^2}}{n \kappa} \ , \notag \\
	\check{x}_n^{AA} &=\frac{-nq+w_0 \mp \sqrt{n^2 +2n q w_0 +w_0^2}}{n \kappa}=\frac{n\kappa}{nq-w_0\pm\sqrt{n^2-2nq w_0 +w_0^2}} \ , \notag \\
	\hat{x}_n^{AS} &=\hat{x}_n^{SA}=\frac{nq+w_0\pm \sqrt{n^2+\Upsilon^2+2nqw_0}}{n \kappa}  \ , \notag \\
	\check{x}_n^{AS} &=\check{x}_n^{SA}=\frac{(n+m) \kappa}{q(n+m)-w_0\pm \sqrt{\Upsilon^2+(m+n)(m+n-2qw_0)}} \ ,
\end{align}
while for the sphere are
\begin{align}
	\kappa \hat{x}_n^{SS} &=\frac{nq+\Upsilon +\sqrt{n^2+2nq\Upsilon +\Upsilon^2}}{n} +\frac{m^2 q \kappa^2}{\Upsilon^2} +\mathcal{O} (\Upsilon^{-3} ) \ , \notag \\
	\kappa \hat{x}_n^{SS} &=\frac{nq+\Upsilon -\sqrt{n^2+2nq\Upsilon +\Upsilon^2}}{n} +\frac{n\kappa^2 \pm \kappa^2 \sqrt{n^2-4m^2}}{2\Upsilon} -\frac{m^2 q \kappa^2}{\Upsilon^2} +\mathcal{O} (\Upsilon^{-3} )  \notag \\
	&= \pm \frac{\kappa^2 \sqrt{n^2-4m^2}}{2\Upsilon}+\frac{(n-2m) q \kappa^2}{2\Upsilon^2} + \mathcal{O} (\Upsilon^{-3} ) \ , \notag \\
	\frac{\kappa}{\check{x}_n^{SS}} &= \frac{q(2m+n)-\Upsilon-\sqrt{(2m+n)^2-2q(2m+n)\Upsilon + \Upsilon^2}}{(2m+n)}+\frac{q \kappa^2 m^2}{\Upsilon^2}+\mathcal{O} (\Upsilon^{-3} )  \ , \notag \\
	\frac{\kappa}{\check{x}_n^{SS}} &= \frac{q(2m+n)-\Upsilon +\sqrt{(2m+n)^2-2q(2m+n)\Upsilon + \Upsilon^2}}{(2m+n)} \notag \\
	&-\frac{(n+2m) \kappa^2 \pm \kappa^2 \sqrt{n(n+4m)}}{2\Upsilon}-\frac{m^2 q \kappa^2}{\Upsilon^2} +\mathcal{O} (\Upsilon^{-3} ) \notag \\
	&= \pm \frac{\kappa^2 \sqrt{n(n+4m)}}{2\Upsilon} +\frac{(n^2 +4mn+2m^2) q\kappa^2}{2\Upsilon^2} +\mathcal{O} (\Upsilon^{-3} ) \ .
\end{align}

Substituting into equation~(\ref{finitegapcorrection}) and using the definition (\ref{definitionOmega}) we get
\begin{align}
	\hat{\Omega}_n^{AA} &=nq+w_0 \pm \sqrt{n^2 +2n q w_0 +w_0^2} \ , \\
	\check{\Omega}_n^{AA} &=nq-w_0\pm\sqrt{n^2-2nq w_0 +w_0^2} \ , \\
	\hat{\Omega}_n^{F} &=nq-\Upsilon \pm \sqrt{n^2+\Upsilon^2+2nqw_0}  \ , \\
	\check{\Omega}_n^{F} &=(m+n)q-w_0\pm \sqrt{\Upsilon^2+(m+n)(m+n-2qw_0)} \ ,
\end{align}
and
\begingroup
\allowdisplaybreaks
\begin{align}
	\hat{\Omega}_n^{SS} &=nq-\Upsilon +\sqrt{n^2+2nq\Upsilon +\Upsilon^2} +\frac{m^2 n q \kappa^2}{\Upsilon^2} + \mathcal{O} (\Upsilon^{-3} ) \notag  \\
	&= 2nq +\frac{n^2 \kappa^2}{2\Upsilon} -\frac{q \kappa^2 n (n^2-2m^2)}{2\Upsilon^2} +\mathcal{O} (\Upsilon^{-3} ) \ , \\
	\hat{\Omega}_n^{SS} &=-2\Upsilon \pm \frac{\kappa^2 n \sqrt{n^2 -4m^2}}{2\Upsilon} +\frac{q \kappa^2 n (n^2 -2m^2)}{2\Upsilon^2} +\mathcal{O} (\Upsilon^{-3} ) \ , \\
	\check{\Omega}_n^{SS} &=q(2m+n) -\Upsilon+\sqrt{(2m+n)^2-2q(2m+n)\Upsilon + \Upsilon^2}+\frac{q \kappa^2 m^2 (2m+n)}{\Upsilon^2}+\mathcal{O} (\Upsilon^{-3} ) \notag \\
	&= -2\Upsilon +2(n+2m) q-\frac{(n+2m)^2 \kappa^2}{2\Upsilon} - \frac{q \kappa^2 (n+2m) [(n+2m)^2 -2m^2]}{2\Upsilon^2} +\mathcal{O} (\Upsilon^{-3} ) \ , \\
	\check{\Omega}_n^{SS} &=\pm \frac{\kappa^2 (n+2m) \sqrt{n (n+4m)}}{2\Upsilon} +\frac{q \kappa^2 (n+2m) (n^2 +4mn+2m^2)}{2\Upsilon^2} +\mathcal{O} (\Upsilon^{-3} ) \notag \\
	&=\pm \frac{\kappa^2 (n+2m) \sqrt{(n+2m)^2 -4m^2}}{2\Upsilon} +\frac{q \kappa^2 (n+2m) [(n+2m)^2 -2m^2 ] }{2\Upsilon^2} +\mathcal{O} (\Upsilon^{-3} )  \ .
\end{align}
\endgroup
We immediately observe that $\hat{\Omega}_n^{AA}$ match frequencies $\omega_{1,n}$ and $\omega_{3,n}$ from equation~(\ref{quadracticfrequenciesads}) up to a shift by a constant, while $\check{\Omega}_n^{AA}$ match frequencies $\omega_{2,n}$ and $\omega_{4,n}$ from equation~(\ref{quadracticfrequenciesads}) up to the opposite shift.

The comparison between fermionic frequencies is not so immediate, since it requires a shift of the mode number $n$. In fact the combination $\hat{\Omega}^F_{n-qw_0}+\frac{w_0}{2}$ is identical to $\omega_{5,n}$ and $\omega_{7,n}$ from the equation~(\ref{Fermionic frequencies}). A similar relation relates $\check{\Omega}_n^F$ with $\omega_{6,n}$ and $\omega_{8,n}$ with an extra contribution from the winding.

The comparison of sphere frequencies is more involved since we do not have a closed expression for general values of $q$. Instead we can compare them at the level of the characteristic equation. From the equation (\ref{Characteristic equations}) and equations (\ref{poleeq1}) and (\ref{poleeq2}) we infer that $\hat{\Omega}^S_n=-\omega^S_n$ and $\check{\Omega}^S_{n-2m}=-\omega^S_n$. Using the reality condition for the fluctuation fields, $\omega^S_n=-\omega^S_{-n}$, these relations can be rewritten as $\hat{\Omega}^S_{-n}=\omega^S_{n}$ and $\check{\Omega}^S_{-n-2m}=\omega^S_n$.


\end{document}

\end{document}